\documentclass[12pt]{article}

\usepackage{fixltx2e}
\usepackage{textcomp}
\usepackage{fullpage}
\usepackage{amsfonts}
\usepackage{verbatim}
\usepackage[english]{babel}
\usepackage{pifont}
\usepackage{color}
\usepackage{setspace}
\usepackage{lscape}
\usepackage{indentfirst}
\usepackage[normalem]{ulem}
\usepackage{booktabs}
\usepackage{natbib}
\usepackage{float}
\usepackage{latexsym}
\usepackage{url}
\usepackage{hyperref}
\usepackage{epsfig}
\usepackage{graphicx}
\usepackage{amssymb}
\usepackage{amsmath}
\usepackage{bm}
\usepackage{array}
\usepackage{mhchem}
\usepackage{ifthen}
\usepackage{caption}
\usepackage{hyperref}
\usepackage{amsthm}
\usepackage{amstext}

\usepackage{graphicx}
\usepackage{multirow}
\usepackage{breqn}
\renewcommand{\tableofcontents}{}

\bibpunct{(}{)}{;}{a}{}{,}  % this is a citation format command for natbib

\begin{document}
%\begin{flushright}
%Version dated: \today
%\end{flushright}
%\bigskip
%\noindent RH:  Relaxed Drift Diffusion
% put in your own RH (running head)
% for POVs the RH is always POINT OF VIEW

\bigskip
\medskip
\begin{center}

% Insert your title:
\noindent{\Large \bf A Relaxed Drift Diffusion Model for Phylogenetic Trait Evolution}
\bigskip

% We don't use a special title page; the author information is entered 
% like any other text.

% FOOTNOTES: We don't allow them in the manuscript, except in
% tables. Don't include any footnotes in the text.

\noindent {\normalsize \sc Mandev S. Gill$^1$, Lam Si Tung Ho$^1$, Guy Baele$^2$, Philippe Lemey$^2$, \\ and Marc A. Suchard$^{1,3,4}$}\\
\noindent {\small \it 
$^1$Department of Biostatistics, Jonathan and Karin Fielding School of Public Health, University
of California, Los Angeles, United States\\
$^2$Department of Microbiology and Immunology, Rega Institute, KU Leuven, Leuven, Belgium \\
$^3$Department of Biomathematics, David Geffen School of Medicine at UCLA, University of California,
  Los Angeles, United States \\
  $^4$Department of Human Genetics, David Geffen School of Medicine at UCLA, Universtiy of California,
  Los Angeles, United States}\\
\end{center}
\medskip

%\begin{center}
 % \setlength{\baselineskip}{0.7cm}
  %{\bf Running head}: Relaxed Drift Diffusion \\ \ \\
  %{\bf Keywords}: Phylogenetics, Phylogeography, Brownian Motion, Phylodynamics, Diffusion Processes, Trait Evolution \\ \ \\
  %{\bf Corresponding Author}: \\
  %Marc A. Suchard \\
  %695 Charles E. Young Dr., South \\
  %Los Angeles, CA 90095-7088 \\
  %Tel: (310) 825-7442 \\
  %Fax:  (310) 825-8685 \\
  %Email: \url{msuchard@ucla.edu}
%\end{center}

%\vspace{1in}
%\clearpage
\begin{abstract}
Understanding the processes that give rise to quantitative measurements associated with molecular sequence data
remains an important issue in statistical phylogenetics.
Examples of such measurements include geographic coordinates in the context of phylogeography
and phenotypic traits in the context of comparative studies.  A popular approach is to model the evolution of
continuously varying traits
as a Brownian diffusion process acting on a phylogenetic tree.  However, standard Brownian diffusion is
quite restrictive and may not accurately characterize certain trait evolutionary processes.  
Here, we relax one of the major restrictions of standard Brownian diffusion by incorporating a nontrivial estimable drift
into the process.  We introduce a relaxed drift diffusion model for the evolution of multivariate continuously
varying traits along a phylogenetic tree via Brownian diffusion with drift.  Notably, the relaxed drift
model accommodates branch-specific variation of drift rates while preserving model identifiability.
Furthermore, our development of a computationally efficient dynamic programming approach to compute
the data likelihood enables scaling of our method to large data sets frequently encountered in viral evolution.
We implement the relaxed drift model in a Bayesian inference framework to simultaneously reconstruct
the evolutionary histories of molecular sequence data and associated multivariate continuous trait data, and
provide tools to visualize evolutionary reconstructions.
We illustrate the utility of our approach in three viral examples.  In the first two, we examine the spatiotemporal spread
of HIV-1 in central Africa and 
West Nile virus in North America and show that a relaxed drift approach uncovers
a clearer, more detailed picture of the dynamics of viral dispersal than standard Brownian diffusion.
Finally, we study antigenic evolution in the context of HIV-1 resistance to three broadly neutralizing antibodies.
Our analysis reveals evidence of a continuous drift 
at the HIV-1 population level
towards enhanced resistance to neutralization by the VRC01 monoclonal antibody over the course of the epidemic.
\end{abstract}

% Points of View do not have abstracts but they should include
% Keywords.

\vspace{1.5in}

\section{Introduction}

Phylogenetic inference has emerged as an important tool for understanding patterns of molecular sequence variation over time.  
%PL: i am not so familiar with the 'nonsequence' term, perhaps rephrase as: ."..a growth of associated data, including spatial and phenotypic data, ..."
%GB: I'm with PL on this ... "a growth of accompanying/associated data/trait data"
Along with the increasing availability of molecular sequence data, there has been a growth of associated sources of information, such as spatial and phenotypic trait data, underscoring the need for integrated models of sequence and trait evolution on phylogenies, which promise to deliver more precise insights %in infectious diseases 
and increase opportunities for statistical hypothesis testing.
\par
Much of the development of trait evolution models has been motivated by phylogenetic comparative approaches focusing on phenotypic and ecological traits.  
%GB: added sentence
Traditional comparative methods assess the correlation between traits through standard regression models that assume taxa traits
are generated independently by the same distribution.  This assumption is obviously violated by taxa traits due to
their shared ancestry.  
%Standard correlations between traits assume independent data, which is obviously violated for taxa traits due to their shared ancestry. 
A proper understanding of patterns of correlation between traits can be achieved only by accounting for their shared evolutionary history \citep{Felsenstein1985, Harvey1991}, and comparative methods focus on relating observed phenotype information to an evolutionary history.
\par
Trait evolution has been tackled from another angle in phylogeographic approaches focusing on geographic locations rather than phenotypic traits.  
Evolutionary change is better understood when accounting for its geographic context, and phylogeographic inference methods aim to connect the evolutionary and spatial history of a population \citep{bloomquist2010three}.  
Phylogeographic techniques have allowed researchers to better understand the origin, spread, and dynamics of emerging infectious diseases.  
Examples include the human influenza A virus \citep{Rambaut2008, Smith2009, Lemey09}, rabies viruses \citep{Biek2007, Seetahal2013}, dengue virus \citep{Bennett2010, allicock2012phylogeography} and hepatitis B virus (e.g. \citet{Mello2013}).
\par
While methods for phenotypic and phylogeographic analyses are developed with different data in mind, they
address similar situations and it is appropriate to speak more generally of trait evolution.  Two key components required
for modeling phylogenetic trait evolution are a method for incorporating phylogenetic information and a model of an evolutionary
process on a phylogeny giving rise to the observed traits.
Many popular approaches first reconstruct a phylogenetic tree and condition inferences pertaining to the trait
evolution process on this fixed tree.
However, computational advances, particularly in Markov chain Monte Carlo (MCMC) sampling techniques,
have made it possible to control for phylogenetic uncertainty (as well as uncertainty in other important model
parameters)
through integrated models that jointly estimate parameters of interest \citep{Huelsenbeck03, Lemey2010}.
\par
The evolution of discrete traits has typically been modeled using 
continuous-time
Markov chains \citep{Felsenstein81, Pagel1999, Lemey2009},
analogous to substitution models for molecular sequence characters.  
However, phenotypic and geographic traits are often continuously distributed,
and while meaningful inferences may still be made partitioning the state space into finite parts, 
stochastic processes with continuous state spaces
represent a more natural approach. %GB: there are also serious problems when using discrete locations; see DeMaio et al. (2015)
A popular choice to model continous trait evolution
along the lineages of a phylogenetic tree is Brownian diffusion \citep{Felsenstein1985}.
\citet{Lemey2010} and \citet{Pybus2012} have recently developed a computationally efficient Brownian diffusion model for evolution of multivariate traits in
a Bayesian framework that integrates it with models for phylogenetic reconstruction and molecular evolution.
Notably, their full probabilistic approach accounts for uncertainty in the phylogeny, demographic history and evolutionary
parameters.  Trait evolution is modeled as a multivariate time-scaled
mixture of Brownian diffusion processes with a zero-mean displacement (or, in other words,
neutral drift) along each branch of the possibly unknown phylogeny.
\par
While adopting a mixture of Brownian diffusion processes is a popular and useful approach, it may not appropriately describe the evolutionary
process in certain situations.  Such scenarios are more realistically modeled by more sophisticated diffusion processes.
There may, for example, be selection toward an optimal trait value.
To address this phenomenon, there has been considerable development of
mean-reverting Ornstein-Uhlenbeck process models for trait evolution,
featuring a stochastic Brownian component along with a deterministic component
%PL: "combining a stochastic Brownian component with a deterministic component"?
\citep{Hansen1997, Butler2004, Bartoszek2012}.
\par
Another trait evolutionary process inadequately modeled by standard Brownian diffusion is
one characterized by directional trends.  The need for relaxing the assumption of neutral
drift is highlighted by a number of evolutionary scenarios in which there are apparent trends
in the direction of variations, including antigenic drift in influenza
\citep{Bedford2014}, the evolution of body mass in carnivores \citep{Lartillot2011}, and dispersal patterns of
viral outbreaks \citep{Pybus2012}.  To this end, we extend the Bayesian multivariate
Brownian diffusion framework of
\citet{Lemey2010} to allow for an unknown estimable nonzero drift vector for the mean displacement
in a computationally efficient manner.
While inclusion of a nontrivial drift represents a promising first step, a constant drift rate
may not hold over an entire evolutionary history.  We address this issue by presenting a flexible
relaxed drift model that permits multiple drift rates on a phylogenetic tree.  Importantly,
we equip the model with machinery to infer the number of different drift rates supported
by the data as well as the locations of rate changes.
\par
We apply our relaxed drift diffusion
methodology to three viral examples of clinical importance.  In the first two examples,
we illustrate our approach in a phylogeographic setting by investigating
the spatial diffusion of HIV-1 in central Africa and the West Nile virus in North America.
For the third example, we explore antigenic evolution
in the context of enhanced resistance of HIV-1 to broadly neutralizing antibodes over the
course of the epidemic.
We employ model
selection techniques to compare the nested drift-neutral, constant drift
and relaxed drift Brownian diffusion models.  We demonstrate a better fit by relaxing the restrictive
drift-neutral assumption, and an improved ability to uncover and quantify key aspects of trait
evolution dynamics.

\section{Methods}

We start by assuming we have a dataset of $N$ aligned molecular sequences $\textbf{X}=(\textbf{X}_1,\dots,\textbf{X}_N)$ along with
$N$ associated $M$-dimensional, continuously varying traits $\textbf{Y}=(\textbf{Y}_1,\dots,\textbf{Y}_N)$.  The sequence and trait data
correspond to the tips of an unknown yet estimable phylogenetic tree $\tau$.  Later we will discuss accounting
for phylogenetic uncertainty, modeling the molecular evolution process giving rise to $\textbf{X}$ and integrating it
with a model for trait evolution.  But first, we explore trait evolution on a fixed phylogeny via a diffusion process
acting conditionally independently along its branches.
\par
%GB: not sure the discussion on the degrees is needed, you already said it's bifurcating
The $N$-tipped bifurcating phylogenetic tree $\tau$ is a graph with a set of vertices
$\mathcal{V} = (\mathcal{V}_1,\dots,\mathcal{V}_{2N-1})$ and edge weights $\mathcal{T}=(t_1,\dots,t_{2N-2})$.  The vertices correspond to nodes of the tree and, as the
length of the tree $\tau$ is measured
in units of time, $\mathcal{T}$ consists of times corresponding to branch lengths.
Each external node $\mathcal{V}_i$ for $i = 1, \dots , N$ is of degree 1, with one
parent node $\mathcal{V}_{pa(i)}$ from within the internal or root nodes.  Each internal node $\mathcal{V}_i$ for $i = N+1, \dots , 2N-2$
is of degree 3 and the root node $\mathcal{V}_{2N-1}$ is of degree 2.  An edge with weight $t_i$ connects $\mathcal{V}_i$ to
$\mathcal{V}_{pa(i)}$, and we refer to this edge as branch $i$.
In addition to the observed traits $\textbf{Y}_1,\dots,\textbf{Y}_N$ at the external nodes, we posit for mathematical convenience
unobserved traits $\textbf{Y}_{N+1},\dots,\textbf{Y}_{2N-1}$ at the internal nodes and root.
\par
Brownian diffusion (also known as a Wiener process) is a continuous-time stochastic process originally developed to
model the random motion of a physical particle \citep{Brown1828, Wiener1958}.  Formally, a standard multivariate Brownian diffusion process
$\textbf{W}(t)$ is characterized by the following properties: $\textbf{W}(0) = \textbf{0}$, the map $t \mapsto \textbf{W}(t)$ is almost surely
%continous,
continuous,
$\textbf{W}(t)$ has
independent increments and, for $0 \leq s \leq t$, $\textbf{W}(t) - \textbf{W}(s)$ has a multivariate normal distribution with mean $\textbf{0}$ and variance
matrix $(t-s) \textbf{I}$.
\par
Recent phylogenetic comparative methods \citep{Vrancken2014,Cybis2015} aim to model the correlated evolution
between multiple traits and, to this end, employ a correlated multivariate Brownian diffusion with displacement variance $(t-s)\textbf{P}^{-1}$.  Here, $\textbf{P}$ is an $M \times M$
infinitesimal precision matrix.  The mean of $\textbf{0}$ posits a neutral drift so that the traits do not evolve according to any systematic directional
trend.  Matrix $\textbf{P}$ determines the intensity and correlation of the trait diffusion after controlling for shared evolutionary history.
\par
The Brownian diffusion process along a phylogeny produces the observed traits by starting at the root node and proceeding down the branches of $\tau$.  The
displacement $\textbf{Y}_i - \textbf{Y}_{pa(i)}$ along a branch is multivariate normally distributed, centered at $\textbf{0}$ with
variance $t_i \textbf{P}^{-1}$ proportional to the length of the branch.  Therefore,
conditioning on the trait value $\textbf{Y}_{pa(i)}$ at the parent node, we have
\begin{equation}
\textbf{Y}_{i} | \textbf{Y}_{pa(i)} \sim N \left(\textbf{Y}_{pa(i)}, t_i \textbf{P}^{-1} \right) .
\end{equation}
An extension that introduces branch-specific mixing parameters $\phi_i$ into the process that rescale
$t_i \mapsto \phi_i t_i$ yields a mixture of Brownian processes and remains popular in phylogeography
\citep{Lemey2010}.

\subsection{Drift}

%GB: why is it nonzero but non-neutral? I have no clue as to these hyphenation rules ...
To incorporate a directional trend, we adopt a multivariate correlated Brownian diffusion process with a non-neutral drift.
In our drift diffusion process we replace the zero mean of the increment $\textbf{W}(t) - \textbf{W}(s)$ with the time-scaled mean
vector $(t-s)\boldsymbol \mu$.
% Notably, the correlation structure of the correlated Brownian diffusion process is unaltered by the incorporation of a non-neutral drift.
The expected difference between the trait values of a descendant and its ancestor is determined
by the overall drift vector $\boldsymbol \mu$ and the time elapsed between descendant and ancestor.
This yields what we will call the constant drift model:
\begin{equation}
\textbf{Y}_i | \textbf{Y}_{pa(i)} \sim N \left( \textbf{Y}_{pa(i)} + t_i \boldsymbol \mu, t_i \textbf{P}^{-1} \right) .
\end{equation}
\par
While this approach is useful for modeling general directional trends, it is quite restrictive in that
the drift $\boldsymbol \mu$ is fixed over the entire phylogeny.  We can relax this assumption by
introducing branch-specific drift vectors $\boldsymbol \mu_i$:
\begin{equation}
\textbf{Y}_i | \textbf{Y}_{pa(i)} \sim N \left( \textbf{Y}_{pa(i)} + t_i \boldsymbol \mu_i, t_i \textbf{P}^{-1} \right)
\end{equation}
for $i = 1, \dots, 2N-2$.
We assign the root a conjugate prior
\begin{equation}
\textbf{Y}_{2N-1} \sim N \left(\boldsymbol \mu^*, (\phi \textbf{P})^{-1} \right),
\end{equation}
that is relatively uninformative for small values of $\phi$. %PL: not sure if it is problematic that $\phi$ has been used earlier in the manuscript as notation for branch-specific scalers in the scaled mixture of normal description.
\par
Conditioning on the trait value $\textbf{Y}_{2N-1}$ at the root of $\tau$,
the joint distribution of observed traits $\textbf{Y}_1,\dots,\textbf{Y}_N$ can be expressed as
\begin{equation}
\text{vec}\left[\textbf{Y} \right] \vert \left(\textbf{Y}_{2N-1}, \textbf{P}, \textbf{V}_{\tau}, \boldsymbol \mu_{\tau} \right)
\sim N \left(\textbf{Y}_{\text{root}} +
\left(\textbf{T} \otimes \textbf{I}_M \right)\boldsymbol \mu_{\tau},
\textbf{P}^{-1} \otimes \textbf{V}_{\tau} \right),
\label{eq:traitLikelihood}
\end{equation}
building on a similar construction for drift-neutral Brownian diffusion \citep{Felsenstein1973, Freckleton2002}.
Here, vec[$\textbf{Y}$] is the vectorization
of the column vectors $\textbf{Y}_1,\dots,\textbf{Y}_N$, %GB: little difficult to read; can we put in an additional 'while' before I_M ?
while $\textbf{I}_M$ is an $M \times M$ identity matrix,
and $\otimes$ is the Kronecker product.
$\textbf{Y}_{\text{root}}$ is the $NM \times 1$
vector $(\textbf{Y}^t_{2N-1},\dots,\textbf{Y}^t_{2N-1})^t$, and
$\boldsymbol \mu_{\tau}$ is the $(2N-2)M \times 1$
drift rate vector $(\boldsymbol \mu^t_1,\dots,\boldsymbol \mu^t_{2N-2})^t$.
The $N \times N$ variance matrix $\textbf{V}_{\tau}$ is a deterministic function of $\tau$
and represents the contribution of the phylogenetic tree to the covariance structure.
Its diagonal entries $V_{ii}$ are equal to the distance in time between the tip $\mathcal{V}_i$ and the
root node $\mathcal{V}_{2N-1}$, and off-diagonal entries $V_{ij}$ correspond to the distance in time between the
root node $\mathcal{V}_{2N-1}$ and the most recent common ancestor of tips $\mathcal{V}_{i}$ and $\mathcal{V}_{j}$.
Finally, the $N \times (2N-2)$ matrix $\textbf T$ is defined as follows: $T_{ij} = t_j$, the length of branch
$j$, if branch $j$ is part of the path from the external node $i$ to the root, and $T_{ij} = 0$ otherwise.
\par
Our development thus far clarifies some important issues.  First, while it is tempting
to model a unique drift rate on each branch,  not all
$\boldsymbol \mu_i$ are uniquely identifiable in the likelihood (5).  Care must be taken to
impose necessary restrictions to ensure identifiability while still permitting sufficient
drift rate variation, and we discuss an approach to achieve this in section 2.3.
Second, the variance matrix $\textbf{P}^{-1} \otimes \textbf{V}_{\tau}$ in (5) suggests
a computational order of $\mathcal{O}(N^3M^3)$ to evaluate the density.  Repeated
evaluation of (5) is necessary for numerical integration in Bayesian modeling,
and viral data sets may encompass thousands of sequences.  Fortunately, \citet{Pybus2012}
demonstrate that phylogenetic Brownian diffusion likelihoods can be evaluated in computational
order $\mathcal{O}(NM^2)$ by modeling %GB: by modeling what in terms of the matrix precision?
in terms of the precision matrix $\textbf{P}$
(as opposed to the variance) and adopting a
dynamic programming approach.  In section 2.2, we present an adaptation of their
algorithm for our drift diffusion likelihood.

\subsection{Multivariate Trait Peeling}

Under our Brownian drift diffusion process,
the joint distribution of all traits is straightforwardly expressed as the product
\begin{equation}
P(\textbf{Y}_1,\dots,\textbf{Y}_{2N-1} | \tau, \textbf{P}, \boldsymbol \mu, \phi) = \left( \prod_{i=1}^{2N-2} P(\textbf{Y}_i|\textbf{Y}_{pa(i)},\textbf{P},t_i,\boldsymbol \mu_i) \right) P(\textbf{Y}_{2N-1} | \textbf{P},\boldsymbol \mu^*, \phi),
\end{equation}
where $\boldsymbol \mu = (\boldsymbol \mu_1,\dots,\boldsymbol \mu_{2N-2},\boldsymbol \mu^*)$.
The density of the observed traits can then be obtained by integrating over all possible realizations of the unobserved traits
at the root and internal nodes.
We adopt a dynamic programming approach that is
analogous to Felsenstein's pruning method \citep{Felsenstein81} and has been employed for drift-neutral
Brownian diffusion likelihoods \citep{Pybus2012, Vrancken2014, Cybis2015} .
\par
We wish to compute the density
\begin{align}
P(\textbf{Y}_1,\dots,\textbf{Y}_{N} )  & =   \int  \cdots \int P(\textbf{Y}_1,\dots,\textbf{Y}_{2N-1}) d \textbf{Y}_{N+1} \dots  d\textbf{Y}_{2N-1} \\
& =  \int \cdots \int  \left( \prod_{i=1}^{2N-2} P(\textbf{Y}_i | \textbf{Y}_{pa(i)} ) \right) P(\textbf{Y}_{2N-1} ) d \textbf{Y}_{N+1} \dots d\textbf{Y}_{2N-1}.
\end{align}
We have omitted dependence on the parameters $\tau, \textbf{P}, \boldsymbol \mu_{1},
\dots, \boldsymbol \mu_{2N-2}, \boldsymbol \mu*$ and $\phi$ from the notation for the sake
of clarity.  The integration proceeds in a postorder traversal integrating out one internal node trait at a time.  Let $\{\textbf{Y}_i \}$ denote the set
of observed trait values descendant from and including the node $V_i$, and suppose $pa(i) = pa(j) = k$.  Our traversal requires computing integrals
of the form
\begin{equation}
P( \{\textbf{Y}_k \} | \textbf{Y}_{pa(k)} ) = \int P(\{\textbf{Y}_i \} | \textbf{Y}_k) P(\{\textbf{Y}_j \} | \textbf{Y}_k) P(\textbf{Y}_k | \textbf{Y}_{pa(k)}) d \textbf{Y}_k .
\end{equation}
Because the integrand is proportional to a multivariate normal density, it suffices to keep track of partial mean vectors $\textbf{m}_k$, partial precision scalars $p_k$
and normalizing constants $\rho_k$.
\par
Let MVN(.;$\boldsymbol \kappa$,$\boldsymbol \Lambda$) denote a multivariate normal probability density function with mean $\boldsymbol \kappa$ and
precision $\boldsymbol \Lambda$.  We can rewrite conditional densities to facilitate integration with respect to the trait at the parent node:
\begin{eqnarray}
P(\textbf{Y}_i | \textbf{Y}_k) & = & \mbox{MVN} \left(\textbf{Y}_i; t_i \boldsymbol \mu_i + \textbf{Y}_{k},
\frac{1}{t_i} \textbf{P} \right) \\
& = & \mbox{MVN} \left(\textbf{Y}_i - t_i \boldsymbol \mu_i; \textbf{Y}_{k},
\frac{1}{t_i} \textbf{P} \right).
\end{eqnarray}
%GB: bit of a strange sentence below
For $i = 1, \dots , N$, set normalizing constant $\rho_i = 1$, partial mean
\begin{equation}
\textbf{m}_i = \textbf{Y}_i - t_i \boldsymbol \mu_i
\end{equation}
and partial precision
\begin{equation}
p_i = \frac{1}{t_i} .
\end{equation}
Then
\begin{eqnarray}
P(\{\textbf{Y}_i \} | \textbf{Y}_k) P(\{\textbf{Y}_j \} | \textbf{Y}_k) & = & \rho_i \rho_j \times \mbox{MVN} (\textbf{m}_i;\textbf{Y}_k,p_i \textbf{P}) \times \\
& & \quad \quad\mbox{MVN}(\textbf{m}_j;\textbf{Y}_k,p_j \textbf{P}) \nonumber \\
& = & \rho_k \times \mbox{MVN}(\textbf{Y}_k; \textbf{m}^*_k, (p_i + p_j) \textbf{P})
\end{eqnarray}
where partial unshifted mean
\begin{equation}
\textbf{m}^*_k = \frac{p_i \textbf{m}_i + p_j \textbf{m}_j}{p_i + p_j} ,
\end{equation}
and normalizing constant
\begin{equation}
\rho_k =  \rho_i \rho_j \left( \frac{p_i p_j}{2 \pi (p_i + p_j)} \right)^{d/2}  |\textbf{P}|^{1/2} \frac {\exp \left[ -\frac{p_i}{2} \textbf{m}_i' \textbf{P} \textbf{m}_i -\frac{p_j}{2}\textbf{m}_j' \textbf{P} \textbf{m}_j  \right] }
{ \exp \left[ -\frac{p_i + p_j}{2} \textbf{m}^{*'}_{k} \textbf{P} \textbf{m}^*_k \right] } .
\end{equation}
Multiplying by $P(\textbf{Y}_k | \textbf{Y}_{pa(k)})$ and integrating with respect to $\textbf{Y}_k$, we get
\begin{eqnarray}
P( \{\textbf{Y}_k \} | \textbf{Y}_{pa(k)} ) & = & \int P(\{\textbf{Y}_i \} | \textbf{Y}_k) P(\{\textbf{Y}_j \} | \textbf{Y}_k) P(\textbf{Y}_k | \textbf{Y}_{pa(k)}) d \textbf{Y}_k \\
& = & \rho_k \times \mbox{MVN}(\textbf{Y}_{pa(k)}; \textbf{m}_k, p_k \textbf{P}),
\end{eqnarray}
where
\begin{equation}
\textbf{m}_k = \textbf{m}^*_k -  t_k \boldsymbol \mu_k,
\end{equation}
and
\begin{equation}
p_k = \frac{1}{  t_k  + \frac{1}{p_i + p_j} } .
\end{equation}
\par
Integrating out all internal node traits yields
\begin{equation}
P(\textbf{Y}_1,\dots,\textbf{Y}_N | \textbf{Y}_{2N-1}) = \rho_{2N-1} \times \mbox{MVN} (\textbf{Y}_{2N-1};\textbf{m}^*_{2N-1},
(p_{2N-2}+p_{2N-3})\textbf{P}) .
\end{equation}
For the
final step, we multiply by the conjugate root prior and integrate:
\begin{eqnarray}
P(\textbf{Y}_1,\dots,\textbf{Y}_N) & = & \int P(\textbf{Y}_1,\dots,\textbf{Y}_N | \textbf{Y}_{2N-1}) P(\textbf{Y}_{2N-1}) d \textbf{Y}_{2N-1} \\
& = &  \rho_{2N-1}  \mbox{MVN} (\textbf{m}^*_{2N-1};\boldsymbol \mu^*, p_{2N-1} \textbf{P}),
\end{eqnarray}
where
\begin{equation}
p_{2N-1} = \frac{(p_{2N-2}+p_{2N-3})\phi}{p_{2N-2}+p_{2N-3} +\phi}.
\end{equation}
%\par
%(Some comments on why this is $\mathcal{O}(NM^2)$)
\par
In practice, the algorithm visits each node in the phylogeny once and computes partial unshifted means $\textbf{m}^*_k$,
partial means $\textbf{m}_k$, partial precisions $p_k$, and normalizing constants $\rho_k$.

\subsection{Identifiability and Relaxed Drift}

Ideally, we would like to model a unique drift rate $\boldsymbol \mu_i$ on each branch $i$ of the
phylogenetic tree.  However, such lax assumptions open the door to misleading inferences.  Adopting
the notation of section 2.1, there can exist distinct drift rate vectors $\boldsymbol \mu_{\tau}
\neq \boldsymbol \mu^*_{\tau}$ such that
\begin{equation}
\left(\textbf{T} \otimes \textbf{I}_M \right)\boldsymbol \mu_{\tau}
= \left(\textbf{T} \otimes \textbf{I}_M \right)\boldsymbol \mu^*_{\tau},
\end{equation}
yielding identical trait likelihoods (\ref{eq:traitLikelihood}).
The lack of model identifiability presents an obstacle to uncovering the ``true''
values of the drift rates that characterize the trait evolution process.
\par
We propose a relaxed drift model that allows for drift rate variation along a phylogenetic
tree while maintaining model identifiability.  This is achieved by having branches inherit
%GB: certain types of rate changes is vague; mention between parentheses that this will be explained in more detail later on?
drift rates from ancestral branches by default, but allowing a random number of specific types
of rate changes to occur along the tree.  We describe the model here and refer readers to the
Appendix for a detailed argument establishing identifiability.
\par
We begin at the unobserved branch leading to
the root, or most recent common ancestor (MRCA), of the phylogenetic tree $\tau$ and
associate with it the drift $\boldsymbol \mu_{\text{\tiny MRCA}}$.  Then the two branches emanating from the
root node either both inherit the drift rate $\boldsymbol \mu_{\text{\tiny MRCA}}$, or a rate change occurs and
one branch receives a new rate while the other branch assumes the rate $\boldsymbol \mu_{\text{\tiny MRCA}}$.
Similarly, whenever a branch splits into two
anywhere in $\tau$, either both child branches assume the same drift
rate as the parent branch, or one child branch takes on a new value while the other
inherits its drift from the parent branch.  Both child branches taking on different drift
rates than the parent branch is not permitted.
\par
Importantly, rather than fix the type of
drift rate transfer that occurs at a given node, we estimate it from the data.
The benefits of this choice
are twofold.  First, a rate change is not forced when the data do not suggest a need for one.
Unnecessarily imposing a large number of unique drift rates to be inferred from limited data
can lead to high variance estimates.  Second, in the event of a rate change occurring at a node,
only one of the two child branches can assume a new drift rate.  We let the data determine which
of the child branches assumes the new rate.  The data may support new rates on both
child branches.
%we are able to infer the relative support for each.
While our model may seem
too restrictive to accommodate such a scenario at first glance, we are able to infer the
relative support for each child, and it is reflected in
the posterior distribution in terms of the probabilities of the two types of changes.
Thus summaries of the posterior distribution can capture the true nature
of drift rate variation in spite of the identifiability restrictions.
\par
It is important to handle the initial drift rate $\boldsymbol \mu_{\text{\tiny MRCA}}$ with care.
One option is to estimate $\boldsymbol \mu_{\text{\tiny MRCA}}$ from the data just as with
all other drift rates.  However, such a choice may not be ideal for data sets that exhibit
relatively long periods of divergence from the MRCA to the sampling times.  There is generally more
information about diffusion dynamics during time periods overlapping with or close to sampling times.
Likewise, the further removed the MRCA is from sampling times, the less information there is about
$\boldsymbol \mu_{\text{\tiny MRCA}}$ and other drift rates near the MRCA.
Because of the interconnectedness of branch drift rates in the relaxed model, estimates of
$\boldsymbol \mu_{\text{\tiny MRCA}}$ and neighboring drift rates under such circumstances
may primarily reflect
information about drift rates on branches near sampling times.  To mitigate misleading inferences of drift
near the MRCA, we can adopt an initial drift $\boldsymbol \mu_{\text{\tiny MRCA}} = 0$ and still interprete changes in drift rates across the tree.
\par
To parameterize the model, we associate a ternary variable $\boldsymbol \delta_k$ with each internal
node $k$ specifying how it passes on its drift rate to its child nodes.
Suppose node $k$ has left child node $i$ and right child node $j$.
If $\boldsymbol \delta_k = -1$, then $\boldsymbol \mu_i = \boldsymbol \mu_k$ and
node $j$ assumes a new rate $\boldsymbol \mu_j = \boldsymbol \mu_k + \boldsymbol \alpha_j$.
If $\boldsymbol \delta_k = 1$, then
node $i$ assumes a new rate $\boldsymbol \mu_i = \boldsymbol \mu_k + \boldsymbol \alpha_i$
while $\boldsymbol \mu_j = \boldsymbol \mu_k$.  If $\boldsymbol \delta_k =0$, then
no drift rate changes occur and $\boldsymbol \mu_k = \boldsymbol \mu_i = \boldsymbol \mu_j$.
To map the ternary $\boldsymbol \delta_k$ variables to binary indicators
$\boldsymbol \gamma_i$ of rate changes
for child branches, we define
\begin{equation}
\boldsymbol \gamma_i = \frac{1 + \boldsymbol \delta_k}{2}|\boldsymbol \delta_k|,
\label{eq:mapA}
\end{equation}
and
\begin{equation}
\boldsymbol \gamma_j = \frac{1 - \boldsymbol \delta_k}{2}|\boldsymbol \delta_k|.
\label{eq:mapB}
\end{equation}
Thus
\begin{equation}
\boldsymbol \mu_i = \boldsymbol \mu_{pa(i)} + \boldsymbol \gamma_i \boldsymbol \alpha_i,
\end{equation}
for $i=1,\dots,2N-2$.  Working with the binary $\boldsymbol \gamma_i$ eases our understanding
of the MCMC procedure to infer drift rate changes, discussed below.  However, we parameterize the
model in terms of the ternary $\boldsymbol \delta_k$ to facilitate enforcement of the model restrictions.
\par
Of particular interest is the random number $K \in {0,\dots,N-1}$ of rate changes that occur
in $\tau$.  We can write $K$ in terms of the $\boldsymbol \delta_i$,
\begin{equation}
	K = \sum_{i=N+1}^{2N-1} |\boldsymbol \delta_i|,
\end{equation}
and it provides us with a natural way to think of the vector
$\boldsymbol \delta = (\boldsymbol \delta_{N+1},\dots,\boldsymbol \delta_{2N-1})$.
For example, we can express our prior beliefs about
$\boldsymbol \delta$ in terms of $K$.  A popular prior for count data is
the Poisson distribution
\begin{equation}
K \sim \mbox{Poisson}(\lambda).
\end{equation}
Here, $\lambda$ is the expected number of rate changes in $\tau$.  In our analyses,
we set $\lambda = \log(2)$, which places
50$\%$ prior probability on the hypothesis of no rate changes.
%In other situations,
%we may not have strong $\it{a}$ $\it{priori}$
%beliefs about $K$, in which case we consider a discrete uniform prior
%\begin{equation}
%K \sim \mbox{Discrete-Uniform}(0,N-1)
%\end{equation}
%that places equal weight on each possible value $K$ can assume.
\par
In order to infer the nature of the drift rate transitions that occur at the
nodes of the phylogenetic tree, we borrow ideas
from Bayesian stochastic search variable selection (BSSVS)
\citep{George1993, Kuo1998, Chipman2001}.
BSSVS is typically applied to model selection problems in a linear regression setting.
In this framework, we begin with a large number $P$ of potential predictors
$\textbf{X}_1,\dots,\textbf{X}_P$ and seek to determine which of them associate
linearly with an $N$-dimensional outcome $\textbf{Y}$.  The full model with all
predictors is
\begin{equation}
\textbf{Y} = \textbf{X}_1 \boldsymbol \beta_1 + \dots + \textbf{X}_P \boldsymbol \beta_P + \boldsymbol \epsilon ,
\end{equation}
where the $\boldsymbol \beta_i$ are regression coefficients and $\boldsymbol \epsilon$ is a vector
of normally distributed error terms with mean $\textbf{0}$.  When a particular
$\boldsymbol \beta_i$ is determined
to differ significantly from 0, the corresponding $\textbf{X}_i$ helps predict $\textbf{Y}$.
If not, $\textbf{X}_i$ contributes little additional information and is fit to be removed from the
model by forcing $\boldsymbol \beta_i = 0$.  Predictors may be highly correlated, and deterministic model selection
strategies tend not to find the optimal set of predictors without exploring all possible subsets.
There exist $2^P$ such subsets, so exploring all of them is computationally unfeasible in general
and fails completely for $P > N$.
\par
BSSVS efficiently explores the possible subsets of model predictors by augmenting the model
state space with a vector $\boldsymbol \delta = (\boldsymbol \delta_1,\dots,\boldsymbol \delta_P)$
of binary indicator variables that dictate which predictors to include.  The indicators
$\boldsymbol \delta_i$ impose a prior on the regression coefficients
$\boldsymbol \beta  = (\boldsymbol \beta_1,\dots,\boldsymbol \beta_P)$ with mean $\textbf{0}$
and variance proportional to a $P \times P$ diagonal matrix with its diagonal equal to
$\boldsymbol \delta$.  If $\boldsymbol \delta_i = 0$, then the prior variance on $\boldsymbol \beta_i$
shrinks to 0 and forces $\boldsymbol \beta_i=0$ in the posterior.  The joint space
$(\boldsymbol \beta, \boldsymbol \delta)$ is explored simultaneously through MCMC.
\par
We apply BSSVS in our relaxed drift setting to determine the types of drift rate transfers that occur.
We achieve this by exploring the joint space $(\boldsymbol \alpha, \boldsymbol \delta)$ of
rate differences between parent and child branches, and ternary rate change indicators.  The
$\boldsymbol \delta_k$ map to binary indicators $\boldsymbol \gamma_i$, as shown in (\ref{eq:mapA}) and
(\ref{eq:mapB}).
We assume that drift rate differences $\boldsymbol \alpha_i = \boldsymbol \mu_i - \boldsymbol \mu_{pa(i)}$
are $\it{a}$ $\it{priori}$ independent and normally distributed,
\begin{equation}
\boldsymbol \alpha_i \sim N( \textbf{0},\boldsymbol \gamma_i \sigma^2 \textbf{I}) .
\end{equation}
If $\boldsymbol \gamma_i = 0$, then the prior variance $\sigma^2$ on the components of $\boldsymbol \alpha_i$ shrinks to
0.  This forces $\boldsymbol \alpha_i = \textbf{0}$, and hence
$\boldsymbol \mu_i = \boldsymbol \mu_{pa(i)}$, in the posterior.
\par
We complete our drift diffusion model specification by assigning
the precision matrix $\textbf{P}$ a Wishart prior with, say, degrees of freedom $v$ and scale
matrix $\textbf{V}$.
Importantly, the Wishart distribution is conjugate to the observed trait likelihood.  Indeed,
invoking the notation for partial means and precisions from section 2.2, the posterior
\begin{equation}
P(\textbf{P} | \textbf{Y}_1,\dots,\textbf{Y}_N) \propto  P(\textbf{Y}_1,\dots,\textbf{Y}_N |\textbf{P}) P(\textbf{P})
\end{equation}
has a Wishart distribution with $N+v$ degrees of freedom and scale matrix
%\begin{equation}
%\left( \textbf{V}^{-1} + p_{2N-1}(\textbf{m}^*_{2N-1} - \boldsymbol \mu^*)(\textbf{m}^*_{2N-1} - \boldsymbol \mu^*)'
%+  \sum_{k=N+1}^{2N-1} [ p_i \textbf{m}_i \textbf{m}'_i + p_j \textbf{m}_j \textbf{m}'_j - (p_i + p_j)\textbf{m}^*_k \textbf{m}^{*'}_k]  \right)^{-1} .
%\end{equation}
\begin{multline}
\Biggl( \textbf{V}^{-1} + p_{2N-1}(\textbf{m}^*_{2N-1} - \boldsymbol \mu^*)(\textbf{m}^*_{2N-1} - \boldsymbol \mu^*)' \\
+ \sum_{k=N+1}^{2N-1} [ p_i \textbf{m}_i \textbf{m}'_i + p_j \textbf{m}_j \textbf{m}'_j - (p_i + p_j)\textbf{m}^*_k \textbf{m}^{*'}_k] \Biggr)^{-1}  .
\end{multline}
\citet{Lemey2010} exploit a similar conjugacy to construct an efficient Gibbs sampler for $\textbf{P}$,
and our adoption of the Wishart prior conveniently allows us to extend use of the sampler to our model
that now includes a relaxed drift process.

\subsection{Joint Modeling and Inference}

A major strength of our Bayesian framework is that it jointly models sequence and trait evolution.
Adopting a standard phylogenetic approach, we
assume the sequence data $\textbf{X}$ arise from a continuous-time Markov chain (CTMC) model for character evolution
acting along the unobserved phylogenetic tree $\tau$.  The CTMC is characterized by a vector $\textbf{Q}$ of mutation parameters that
may include, for instance, relative exchange rates among characters, an overall rate multiplier and across-site variation
specifications.
The traits $\textbf{Y}$ arise from a Brownian drift diffusion process acting on $\tau$, governed by parameters $\boldsymbol \Lambda$.  A crucial
%GB: should we downplay this w.r.t. the structured coalescent, where this independence does not hold?
assumption is that the processes giving rise to the observed sequences and traits are
conditionally independent given the phylogenetic tree $\tau$:
\begin{equation}
P(\textbf{X},\textbf{Y}|\tau,\textbf{Q},\boldsymbol \Lambda) = P(\textbf{X}|\tau, \textbf{Q})P(\textbf{Y}|\tau, \boldsymbol \Lambda),
\end{equation}
enabling us to write the joint model posterior distribution as
\begin{eqnarray}
P(\tau, \textbf{Q}, \boldsymbol \Lambda | \textbf{X},\textbf{Y}) \propto P(\textbf{X}|\tau,\textbf{Q}) P(\textbf{Y}|\tau,\boldsymbol \Lambda)
P(\tau)P(\textbf{Q})P(\boldsymbol \Lambda) .
\label{eq:fullJointPosterior}
\end{eqnarray}
\par
We implement the joint model by integrating our drift diffusion framework for trait evolution into the
Bayesian Evolutionary Analysis Sampling Trees (BEAST) software
package \citep{Drummond2012}.  BEAST provides an array of
efficient methods for Bayesian phylogenetic inference,
particularly to estimate phylogenies and model molecular sequence evolution.  For the phylogeny $\tau$, we
choose from flexible coalescent-based priors that do not make strong $\it{a}$ $\it{priori}$ assumptions
about the population history \citep{Minin2008, Gill2013}.  For sequence evolution,
we have access to a range of classic substitution models \citep{Kimura1980, Felsenstein81, Hasegawa1985},
gamma-distributed rate heterogeneity among sites \citep{Yang1994}, and strict and relaxed molecular
clock models for branch rates \citep{Drummond2006}.
\par
Estimation of the full joint posterior (\ref{eq:fullJointPosterior}) is achieved through MCMC sampling
\citep{Metropolis1953, Hastings1970}.  We employ standard Metropolis-Hastings transition
kernels available in BEAST to integrate over the parameter spaces of $\textbf{Q}$ and $\tau$.
To sample realizations of the drift diffusion precision matrix $\textbf{P}$, we adapt a Gibbs sampler
developed for drift-neutral Brownian diffusion \citep{Lemey2010}.
For the relaxed drift model, we need transition kernels to explore the space
$(\boldsymbol \alpha, \boldsymbol \delta)$ of branch rate differences and ternary rate
change indicators.  We propose new rate differences $\boldsymbol \alpha^*_i$ component-wise
through a random walk transition kernel that adds random values within a specified window size to the
current $\boldsymbol \alpha_i$.
\par
For $\boldsymbol \delta$, we implement a
trit-flip transition kernel that chooses one of the $N-1$ ternary indicators $\boldsymbol \delta_k$
uniformly at random and proposes a new state $\boldsymbol \delta^*_k$ assuming one of the two possible
values not equal to $\boldsymbol \delta_k$ with equal probability.  For example, if
$\boldsymbol \delta_k = 0$, then
\begin{equation}
\boldsymbol \delta^*_k =
\begin{cases}
-1 & \text{with probability } \frac{1}{2} \\
1 & \text{with probability } \frac{1}{2} .
\end{cases}
\end{equation}
%At first glance, such a transition kernel density suggests a symmetric proposal ratio
%$q(\boldsymbol \delta | \boldsymbol \delta^*)/q(\boldsymbol \delta^* | \boldsymbol \delta) = 1$.
%However,
We have parameterized our prior on $\boldsymbol \delta$ in terms of the
number $K$ of rate changes, and this parameterization should be retained for the transition
kernel in order to ensure the correct Metropolis-Hastings proposal ratio \citep{Drummond2010}.
A proposed increase in rate changes occurs when we choose a $\boldsymbol \delta_k$ with
value 0, so
\begin{equation}
q(K^*=K+1 | K) = \frac{N-1-K}{N-1} .
\end{equation}
If we choose a $\boldsymbol \delta_k$ with a nonzero value, we propose the other nonzero value
with probability 0.5 (corresponding to $K^*=K$), and we propose 0 with probability 0.5
(which means a decrease in rate changes, $K^*=K-1$).  Therefore
\begin{equation}
q(K^*=K|K) = q(K^*=K-1|K) = \frac{1}{2}\frac{K}{N-1}.
\end{equation}
These calculations yield the following proposal ratio for $K$:
\begin{equation}
\frac{q(K|K^*)}{q(K^*|K)} =
\begin{cases}
\frac{1}{2} \frac{K+1}{N-1-K} & \text{if } K^* = K+1 \\
1 & \text{if } K^* = K \\
\frac{2(N-K)}{K} & \text{if } K^* = K-1 .
\end{cases}
\end{equation}
\par
In addition to the parameters characterizing the trait and sequence evolution processes,
we may wish to make inferences about the posterior distribution of traits at the root and
internal nodes, or at any arbitrary time point in the past.
We equip BEAST with the ability to generate posterior trait realizations at these
nodes by implementing a preorder, tree-traversal algorithm.
\par
A natural choice to summarize the results of a Bayesian phylogenetic analysis is a
maximum clade credibility (MCC) tree.  To form  an MCC tree, the posterior sample of trees
is examined to determine posterior clade probabilities, and the tree with the maximum
product of posterior clade probabilities is the MCC tree.  The branches and nodes of MCC
trees can be annotated with inferred drift rates and trait values, along with other
quantities of interest.
Annotated MCC trees can be %constructed with TreeAnnotator and
summarised using TreeAnnotator, available as part of the BEAST distribution, and %PL: because technically Figtree is no part of the BEAST distribution
visualized using FigTree (\url{http://tree.bio.ed.ac.uk/software/figtree/}).

\subsection{Model Selection}

We can formally compare the constant and relaxed drift diffusion models through
Bayes factors (BFs).  A Bayes factor \citep{Jeffreys1935,Jeffreys1961} compares
the fit of two models, $M_1$ and $M_0$, to observed data $(\textbf{X},\textbf{Y})$ by taking the
ratio of marginal likelihoods:
\begin{equation}
BF_{10} = \frac{P( \textbf{X},\textbf{Y} |M_1)}{P( \textbf{X},\textbf{Y} | M_0) } =
\frac{P(M_1 | \textbf{X},\textbf{Y} )}{P(M_0 | \textbf{X},\textbf{Y} ) } \left/ \frac{P(M_1)}{P(M_0)} \right. .
\end{equation}
$BF_{10}$ quantifies the evidence in favor of model $M_1$ over $M_0$.  \citet{Kass1995}
provide guidelines for assessing the strength of the evidence against $M_0$: BFs
between 1 and 3 are not worth more than a bare mention, while values between 3 and 20
are considered positive evidence against $M_0$.  BFs in the ranges 20-150 and $>$150
are considered to be strong and very strong evidence against $M_0$, respectively.
\par
Evaluation of Bayes factors has become a popular approach to model selection in Bayesian
phylogenetics \citep{Sinsheimer1996, Suchard2001, Suchard2005}.  Marginal likelihood estimation can be quite
difficult in a phylogenetic context, and
stepping-stone sampling estimators have been implemented to address this
\citep{Baele2012, Baele2012b, Baele2013}.  Following the approach of \citet{Drummond2010}, however,
we are able to straightforwardly compute the Bayes factor $BF_C$ supporting the constant
drift model $M_C$ over the relaxed drift model $M_R$.  The model $M_C$ is nested within the more
general model $M_R$ and occurs when $K=0$.  This enables us to write
\begin{eqnarray}
BF_{C} & = & \frac{P(\textbf{X},\textbf{Y} |M_C)}{P(\textbf{X}, \textbf{Y} | M_R) } =
\frac{P(M_C | \textbf{X}, \textbf{Y})}{P(M_R | \textbf{X}, \textbf{Y}) }
\left/ \frac{P(M_C)}{P(M_R)} \right. \\[2ex]
& = & \frac{P(K=0 | \textbf{X}, \textbf{Y}, M_R)}{1-P(K=0 | \textbf{X}, \textbf{Y}, M_R) }
\left/ \frac{P(K=0 | M_R)}{1-P(K=0|M_R)} \right. ,
\end{eqnarray}
requiring only our prior probability of no rate changes under the relaxed drift model, and
the posterior probability of zero rate changes.

\section{The Spread of HIV-1 in Central Africa}

\citet{Faria2014} explore the early spatial expansion and epidemic dynamics of HIV-1 in central Africa by
analyzing sequence data sampled from countries in the Congo River basin.  The authors employ
a discrete phylogeographic inference framework \citep{Lemey2009} and show that
the pandemic likely originated in Kinshasa (in what is now the Democratic Republic of Congo) in the
1920s.  Furthermore, viral spread to other population centers in sub-Saharan Africa
was aided by a combination of factors, including strong railway networks, urban growth, and
changes in sexual behavior.
\par
%GB: so the original paper used discrete phylogeography rather than continuous? should be mentioned more explicitly?
We follow up the analysis of \citet{Faria2014}
by applying
our continuous drift diffusion approach
to one of the data sets analysed in this study.
The data set consists of HIV-1 sequences sampled between 1985-2004 from the Democratic
Republic of Congo and the Republic of Congo
%.  The data include
and includes
96 sequences from Kinshasa
\citep{Kalish2004, Vidal2000, Vidal2005, Yang2005},
96 sequences from Mbuji-Mayi \citep{Vidal2000, Vidal2005},
96 from Brazzaville \citep{Bikandou2004, Niama2006},
76 from Lubumbashi \citep{Vidal2005},
33 from Bwamanda \citep{Vidal2000},
24 from Likasi \citep{Kita2004},
23 from Kisangani \citep{Vidal2005},
and 22 sequences from Pointe-Noire \citep{Bikandou2000}.
We reconstruct the spatial dynamics under drift-neutral, constant drift, and relaxed
drift Brownian diffusion
on a maximum clade credibility tree estimated from the sequences and their locations of sampling.
The traits in this instance
are bivariate longitude and latitude coordinates, with observed traits corresponding to
sampling locations.  Table 1 reports posterior estimates of drift.

\begin{table}[htb]
\begin{center}
\begin{tabular}{lcccccc}
%\multicolumn{5}{c}{Table 1: Spatiotemporal Dynamics of HIV-1 in Central Africa} \\
\cline{1-7}
 & \multicolumn{2}{c}{ No Drift} & \multicolumn{2}{c}{ Constant Drift} & \multicolumn{2}{c}{ Relaxed Drift} \\
%	& No Drift  & Constant Drift & Relaxed Drift \\
\hline
Drift (Lat.) & - & - & -0.09 & (-0.11, -0.06)  &  -0.03 & (-0.05, -0.01) \\
Rate Changes (Lat.) & - & - & - & - & 28.13 & (27.0, 29.0) \\
Variance (Lat.) & 0.25 & (0.23, 0.27) & 0.23 & (0.21, 0.25) & 0.13 & (0.12, 0.14) \\
Drift (Long.) & - & - & 0.30 & (0.26, 0.33) & 0.12 & (0.08, 0.16)   \\
Rate Changes (Long.) & - & - & - & - & 1.48 & (1.00, 3.00) \\
Variance (Long.) & 0.59 & (0.55, 0.64) & 0.37 & (0.34, 0.40) &  0.43 & (0.39, 0.45) \\
Correlation & -0.47 & (-0.54, -0.40) & -0.40 & (-0.47, -0.32) & -0.84 & (-0.87, -0.82) \\
%Dispersal Rate (km/year) & & & \\
\hline
\end{tabular}
\end{center}
\caption[Spatiotemporal dynamics of HIV-1 in central Africa]
{Spatiotemporal dynamics of HIV-1 in central Africa.  Model comparison of
Brownian diffusion with no drift, constant drift Brownian diffusion, and relaxed drift
Brownian diffusion.  We report posterior mean estimates along with 95$\%$ Bayesian
credibility intervals (BCIs).
Drift rates for latitude and longitude coordinates
are reported in units of degrees per year.
\label{tb:hivcongo}}
\end{table}

%\clearpage

\par
Under the constant drift model, we infer a significant longitudinal drift
with posterior mean 0.30 degrees per year and a 95$\%$ Bayesian credibility interval (BCI) of (0.26, 0.33), as well as a significant latitudinal drift
with posterior mean of -0.09 degrees per year and BCI (-0.11, -0.06).  Furthermore, for
each coordinate the Bayes factor in favor
of a constant drift model over a drift-neutral model is greater than 1,000, indicating
a substantially better fit for the constant drift model.  These results imply general
eastward and southward trends in the spread of HIV from the Kinshasa-Brazzaville-Pointe-Noire
area to other population centers.  They also reflect the composition of sampling locations: in terms of
longitude, a majority are far to the east of the believed origin while the rest are relatively close to it.
Similarly, nearly 90$\%$ of the sequences come from locations south of Kinshasa or from neighboring locations
of similar latitude.  On the other hand, the existence of samples from cities north of Kinshasa, Bwamanda and Kisangani,
suggests that diffusion with a northward trend may more accurately characterize part of the spatial history.
We explore this possibility with the relaxed drift model.
\par
Under the relaxed drift model, there is significant evidence of multiple longitudinal drift rates.
We estimate a posterior mean of 1.48 drift rate changes with BCI (1, 3), and the Bayes factor in favor of
relaxed drift over constant drift is greater than 1,000.  The posterior mean longitudinal drift rate across all branches
is 0.12 with BCI (0.08, 0.16).  Figure 1 shows the maximum clade credibility tree colored according to
drift rates.  The tree is essentially divided up into two clades: one clade with green-colored branches
and another with brown-colored branches.  We infer eastward drift rates of 0.18
degrees per year on green branches, and drift rates close to or equal to 0 on brown branches.  Tree nodes in Figure 1
are depicted as circles of different sizes, with the size of each circle being determined by the
longitude of the observed or inferred location corresponding to the node.  Larger circles represent
more eastward locations.  The observed and inferred longitudes provide better understanding of the difference
in drift rates between the two clades.
During the period 1960-2004, there is generally greater eastward movement along the lineages
of the green clade.  Although the lineages of the brown clade show greater eastward spread during
the first half of the evolutionary history, the drift rates in the two clades are driven 
by the trends
of the second half of the evolutionary history.  The second half accounts for a much
greater proportion of tree branches and, because it overlaps with all sampling times, contains more information
about the spatial diffusion process.
%The inferred longitudes suggest that the spread during the first half of
%the evolutionary history may be more accurately characterized by the brown clade branches
%exhibiting a greater eastward drift than green clade branches.

\begin{figure}
	\centering
	 \includegraphics[width=1.0\textwidth]{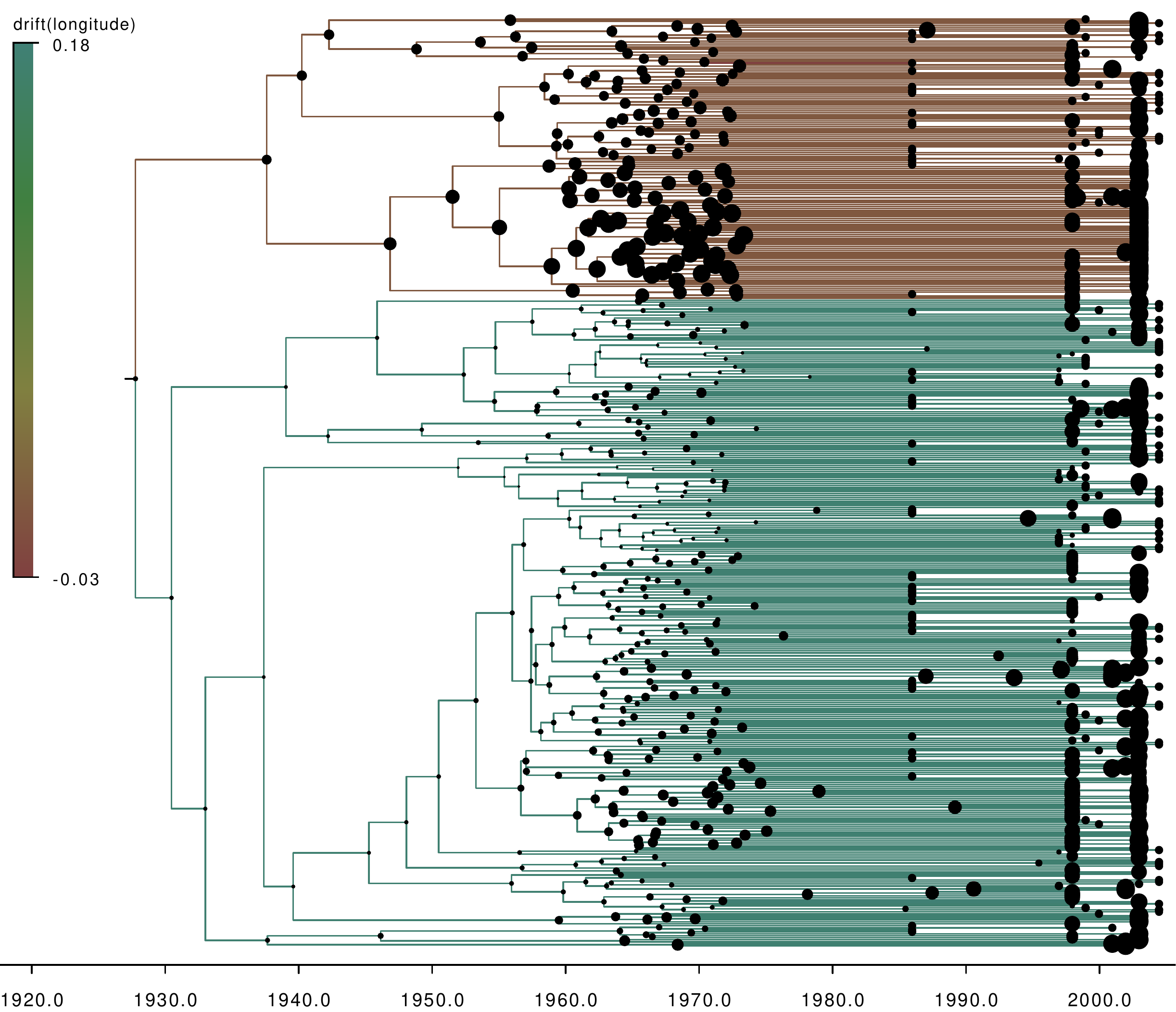}
 	\caption[Maximum clade credibility tree for spread of HIV-1 in central Africa
 	with East-West spatiotemporal dynamics]
 	{Maximum clade credibility tree for spread of HIV-1 in central Africa.
 	The posterior mean longitudinal drift
 	is depicted using a color gradient along the branches.  Green indicates an eastward
 	drift while brown signifies drift rates close to or equal to zero.
	Tree nodes are depicted as circles of different sizes.  The size of each circle is determined
	by the longitude of the observed or inferred location corresponding to the node.  Larger
	circles represent more eastward locations.
	}
 	\label{Fig. 1}
\end{figure}

\begin{figure}
	\centering
	 \includegraphics[width=1.0\textwidth]{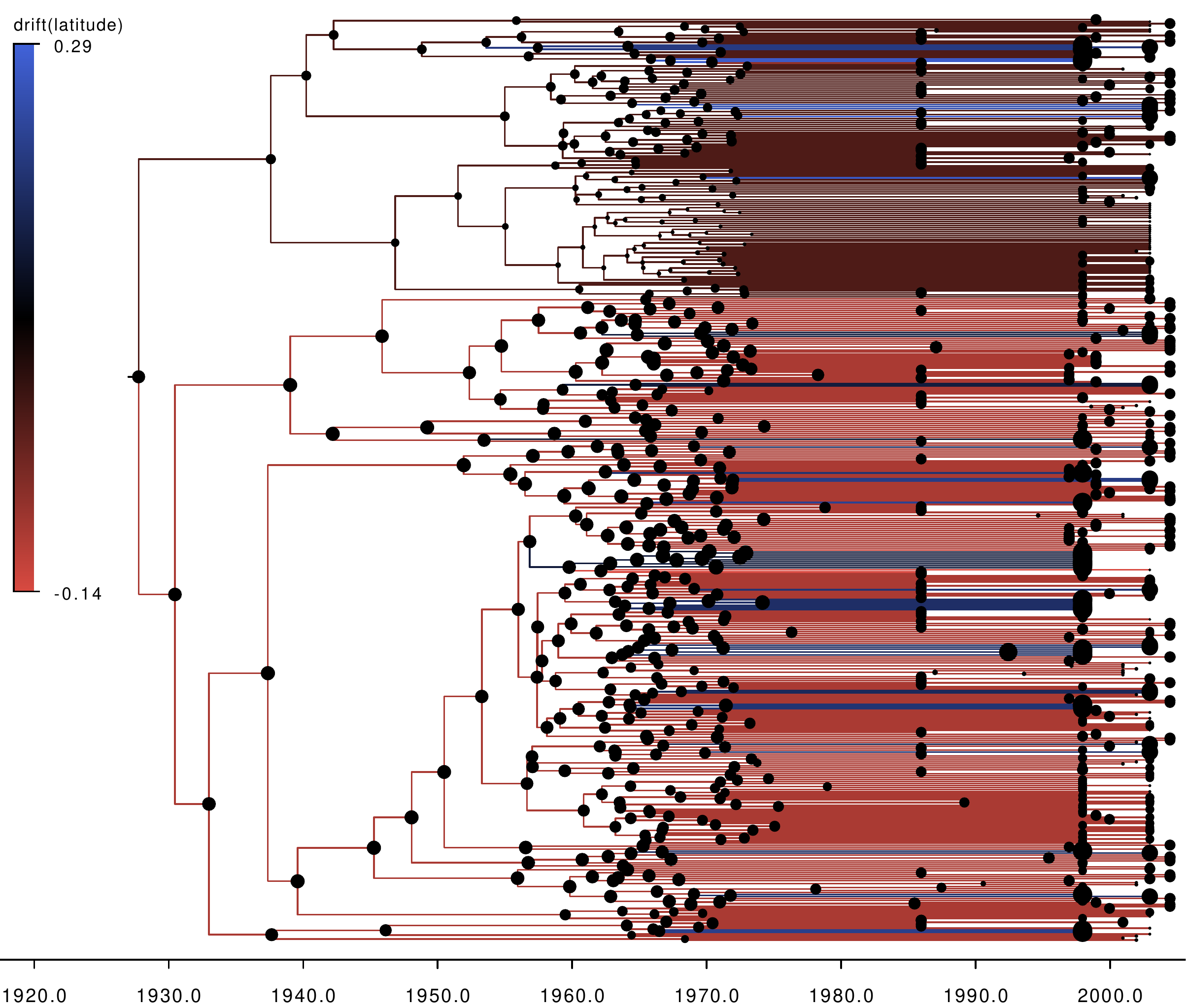}
 	\caption[Maximum clade credibility tree for spread of HIV-1 in central Africa with 
 	North-South spatiotemporal dynamics]
 	{Maximum clade credibility tree for spread of HIV-1 in central Africa.
 	The posterior mean latitudinal drift
 	is depicted using a color gradient along the branches.  The colors range between
 	red and blue, with the former indicating a southward drift and
 	the latter a northward drift.  Tree nodes are depicted as circles of different sizes.
 	The size of each circle is determined
	by the longitude of the observed or inferred location corresponding to the node.  Larger
	circles represent more northward locations.}
 	\label{Fig. 2}
\end{figure}

\par
For latitudinal drift, we estimate 28.13 rate changes with BCI (27, 29).
Furthermore, relaxed drift is supported over constant drift with a Bayes factor greater than 
1,000.  The overall posterior mean drift rate across all branches is -0.03 with BCI (-0.05, -0.01).
Figure 2 depicts an annotated maximum clade credibility tree colored according to latitudinal drift 
rates.  The
color gradient ranges from red to black to blue, with red indicating a southward drift, blue a northward
drift, and black a neutral drift.  As in Figure 1, tree nodes are depicted as circles of varying sizes,
with larger circles representing greater observed or inferred latitude coordinates.
Notably, external branches with positive drift rates lead to samples from locations north of the
origin (Bwamanda and Kisangani), while external branches with nonpositive drift rates lead to
samples from locations with latitudes south of or similar to the origin.
\par
Importantly, adopting a zero-mean displacement distribution when the diffusion process is
more accurately described with a nontrivial mean can result in inflated displacement 
variance rate estimates.  (Recall that the displacement variance along a branch of length
$t_i$ is $t_i \textbf P^{-1}$, so by ``displacement variance rates,'' we mean the diagonal
elements of the variance rate matrix $\textbf P^{-1}$.) 
By incorporating drift into the model, we are able to
disentangle the drift from the variance and uncover a clearer picture of the movement.
The variance rate estimates in Table 1 illustrate this point.
Including a constant drift reduces the variance rate of the displacement of the longitude coordinate from
0.59 to 0.37.  For the latitude, on the other hand,
% going from no drift to a constant drift does not appreciably reduce the variance of the displacement.
the variance rate decreases modestly from 0.25 to 0.23 and the BCIs
of (0.23, 0.27) and (0.21, 0.25) overlap.  The lack of an appreciable reduction in variance rate 
%when going from no drift to a constant drift
may be explained by the fact that the apparent northward drift to Bwamanda and Kisangani remains
unaccounted for by inclusion of a constant southward drift rate.  Indeed, by
accommodating drift rate changes under the relaxed drift
model, the variance rate of the latitudinal displacement drops to 0.13 with BCI (0.12, 0.14).

\section{West Nile Virus}

%The West Nile virus (WNV) is an arbovirus spread primarily by infected mosquitos.
%GB: this isn't the first time you mention WNV so the abbreviation should be introduced at first mention
West Nile virus (WNV) is the most important mosquito-transmitted arbovirus now native to the U.S.
Birds are the most common host,
although WNV has been known to infect other animals including humans.  About 70-80$\%$ of infected humans do not
develop any symptoms, and most of the remaining infections result in fever and other symptoms such as headaches,
body aches, joint pains, vomiting, diarrhea or rash.  Fewer than 1$\%$ of infected humans will develop a serious
neurologic illness such as encephalitis or meningitis.  However, recovery from neurologic infection can take several months,
some neurologic effects can be permanent, and about 10$\%$ of such infections lead to death.
%PL: not sure the clinical aspects need to be so detailed as in the above
Since 1999, WNV has been responsible for more than 1,500 deaths in the United States \citep{CDC2013}. %PL: in the U.S.?
\par
WNV was first detected in the United States in New York City in August 1999, and is most closely
related to a highly pathogenic WNV lineage isolated in Israel in 1998 \citep{Lanciotti1999}.
Surveillance records of WNV incidence show a wave of infection that spread westward and reached the west
coast by 2004 \citep{CDC2013}.  Thus the spread of WNV in North America can be naturally modeled as a spatial
diffusion process with a directional trend.
\par
We analyze a data set of 104 WNV complete genome sequences sampled between 1999 and 2008 and isolated from a
number of different host and vector species \citep{Pybus2012}.  The sequences were sampled from a wide variety
of locations in the United States and Mexico.  We represent the sampling locations as bivariate traits consisting
of latitude and longitude coordinates and conduct a phylogeographic analysis using our Brownian drift diffusion model.
Results are presented in Table 2.

\begin{table}[htb]
\begin{center}
\begin{tabular}{lcccc}
%\multicolumn{5}{c}{Table 2: Spatiotemporal Dynamics of WNV in North America} \\
\cline{1-5}
& \multicolumn{2}{c}{ No Drift} & \multicolumn{2}{c}{Constant Drift} \\
	% & No Drift  & Constant Drift \\
\hline
Drift (Lat.) & - & - & -0.25 & (-1.05, 0.08)  \\
Variance (Lat.) & 6.74 & (4.83, 15.81) & 6.38 & (4.60, 14.61) \\
Drift (Long.) & - & - & -1.77 & (-3.43, -0.27)    \\
Variance (Long.) & 23.37 & (16.95, 54.04) & 20.06 & (14.67, 45.52) \\
Correlation & 0.22 & (0.03, 0.43) & 0.18 & (-0.03, 0.36)  \\
%Dispersal Rate (km/year) & & & \\
\hline
\end{tabular}
\end{center}
\caption[Spatiotemporal dynamics of West Nile virus in North America]
{Spatiotemporal dynamics of West Nile virus in North America.  Model comparison of
Brownian diffusion with no drift and constant drift Brownian diffusion.
We report posterior mean estimates along with 95$\%$ BCIs.
Drift rates for latitude and longitude coordinates
are reported in units of degrees per year.
\label{tb:wnv}}
\end{table}

\par
We begin by analyzing the data with the constant drift model.  For the longitude, the posterior mean drift rate is
-1.77 degrees per year with a 95$\%$ BCI of (-3.43, -0.27).
This indicates a significant westward trend in the spread of WNV, consistent with WNV incidence data.
Furthermore, a Bayes factor of 18.23 lends considerable support to the constant drift model over the
drift-neutral model.
For the latitude, we have an estimated posterior mean drift of -0.25 degrees per year, but the
95$\%$ BCI of (-1.05, 0.08) contains zero.  Moreover, the Bayes factor in favor of constant drift over no drift
is 0.71.  Therefore there is not much evidence of a significant North-South drift.  We check for the possibility
of multiple drift rates under the relaxed drift model.  However, the inferred number of rate changes in the
latitudinal drift is 0.63 with BCI (0, 2), and the Bayes factor in favor of relaxed drift over constant drift
is 0.87.  Similarly, for the longitude coordinate, we estimate 0.71 rate changes with BCI
(0, 2), and Bayes factor of 1.04.  Thus the data do not support relaxed drift over
constant drift.
\par
While inclusion of drift leads to smaller displacement variance rates in our analysis of HIV-1 dispersal
dynamics, Table 2 shows that not to be the case for the WNV data.  Even for the longitude coordinate
where there is significant evidence of a nontrivial drift, there is a great deal of BCI overlap in
the estimated displacement variance rate for drift-neutral and constant drift diffusion.
This difference can be explained by comparing the magnitude of the displacement drift relative
to the displacement standard deviation in the WNV and HIV examples.  Here, we work with
displacement standard deviation rather than displacement variance in order to make comparisons
on the same scale.  Recall that the displacement mean and variance along a branch 
are both proportional to the branch length.  To get an estimate of the average displacement mean
and standard deviation for each example, we use the mean branch length.
The mean branch length is 16.9 years for the HIV data set, and 2.4 years for the WNV
data set.  Under constant drift, we get an average longitudinal 
displacement mean of 5.07 for the HIV data, and -4.25 for the WNV data.  
The average longitudinal displacement standard deviation without drift is 3.25 for the HIV data,
and 7.49 for the WNV data.
%For the HIV
%data, the average displacement standard deviation is 3.15 without drift, and 
%2.50 with a constant drift.  The 
%corresponding estimates are 7.49 and 6.94, respectively, in the case of WNV. 
In the HIV analysis, the average displacement mean with drift is
1.61 times as much as the average displacement standard deviation without drift.
In the case of the WNV, the absolute value of the 
average displacement mean with drift is
0.57 times as much as the average displacement standard deviation without drift.
%standard deviation without drift is about 
%0.62 times as much as the average displacement mean with drift.  
%On the other hand, 
%in the case of WNV, the average displacement standard deviation without drift is about 
%1.76 times the absolute value of the displacement mean with drift.
So while the displacement standard deviation in the drift-neutral HIV analysis is inflated by the
hidden drift, the latent drift represents a much smaller contribution
to the drift-neutral displacement standard deviation in the WNV analysis.

\section{HIV-1 Resistance to Broadly Neutralizing Antibodies}

It is widely believed that a successful HIV-1 vaccine will require the elicitation of
neutralizing antibodies \citep{Johnston2007, Barouch2008, Walker2008}.  Most neutralizing antibodies are
strain-specific and therefore not so attractive for vaccine design \citep{Weiss1985, Mascola2010}.
It is important to identify and characterize antibody specificities that are effective against
a large number of currently circulating HIV-1 variants \citep{Burton2002, Burton2004}.  Several
broadly neutralizing monoclonal antibodies have been recently isolated, including
PG9 and PG16 \citep{Walker2009}, and VRC01 \citep{Zhou2010}.
\par
Studies comparing viruses isolated from individuals who seroconverted early in the HIV-1 epidemic to viruses
from individuals who seroconverted in recent years have shown that HIV-1 has become increasingly resistant to
antibody neutralization over the course of the epidemic \citep{Bunnik2010,Euler2011,Bouvin-Pley2013}.
\citet{Bunnik2010} demonstrate a decreased sensitivity to polyclonal antibodies and
to monoclonal antibody b12.  \citet{Euler2011} extend those findings by investigating
whether HIV-1 has adapted to the neutralization activity of PG9, PG16, and VRC01.  Their results
show that HIV-1 has become significantly more resistant to neutralization by VRC01 and also provide
some support for increased resistance to neutralization by PG16.
\par
These studies typically do not account for phylogenetic dependence among the sampled viruses.
\citet{Vrancken2014} examine the data set of \citet{Euler2011} with a
Brownian diffusion trait evolution model that simultaneously infers phylogenetic signal,
the degree to which resemblance in traits reflects phylogenetic relatedness.
They find moderate phylogenetic signal and, through ancestral trait value reconstruction, more
evidence of decreased sensitivity of HIV-1 to VRC01 and PG16 neutralization at the population level.
\par
We follow up on the analysis of \citet{Vrancken2014} by incorporating drift into the Brownian
trait evolution.
%We apply our drift diffusion model to the data set of to study the
%population-level evolution of resistance of
%HIV-1 to broadly neutralizing antibodies PG9, PG16 and VRC01.
The data set is comprised of
clonal HIV-1 variants
from ``historic'' and ``contemporary'' seroconverters with an acute or early subtype B HIV-1 infection.
The 14 historic seroconverters have a known seroconversion date between 1985 and 1989, and the 21
contemporary seroconverters have a seroconversion date between 2003 and 2006.  The percent neutralization
is determined by calculating the reduction in p24 production in the presence of the neutralizing agent
compared to the p24 levels in the cultures with virus only.  The trait values of interest
are 50$\%$ inhibitory concentration (IC$_{50}$) assay values that summarize the percent neutralization
by antibodies PG9, PG16 and VRC01, measured in units of $\mu$g/ml.
%GB: not sure about the following sentence ...
We take the log-transform of IC$_{50}$ values in order to ensure that concentration values are
strictly positive under the diffusion process.  Higher $\log($IC$_{50})$ values
correspond to greater resistance to antibody neutralization.
For viruses with $\log($IC$_{50})$ values that fall outside
the tested antibody concentration range, we integrate out the concentration over a plausible IC$_{50}$
interval.
\par
First, we analyze the data with the constant drift model (see Table 3).  The results are
essentially consistent with the
findings of \citet{Euler2011}.  For VRC01, we estimate a posterior
mean drift of 0.15 with 95$\%$ BCI (0.06, 0.24), signaling a significant drift toward
higher resistance to VRC01 neutralization.  Furthermore, a Bayes factor of 32.33
lends strong support to a constant drift over no drift.
There is not as much evidence of a trend for PG16.
On one hand, the posterior probability
that the drift rate is positive is 0.953, providing some corroboration for a decreased sensitivity to
PG16 neutralization.  However, we infer a mean drift rate of 0.10
with a 95$\%$ BCI of (-0.02, 0.20) that contains zero.  Furthermore, the Bayes factor in favor
of constant drift over no drift is just 1.32, showing little support for inclusion of a drift term.
We do not detect a significant drift in the case of PG9: the posterior mean is 0.08 with
95$\%$ BCI (-0.05, 0.19) and the Bayes factor is 1.17.
% and the probability of a positive drift rate is less than 0.9.
\par
To take a closer look, we fit the data to the relaxed drift model.
% We assign a Poisson prior to the number of rate changes, with $\lambda = \log(2)$, so that there's 50$\%$ prior probability on the hypothesis of no rate changes.
Along with branch specific drift rates, we examine
the posterior mean rates over the entire evolutionary history (see Table 3).
%The posterior mean drift rates are similar to the estimates from the constant drift model and are presented in Table 3.
First, we consider the results for PG9 and PG16.  The posterior mean drift rates are similar to the inferred
drift rates under the constant drift model,
and their 95$\%$ BCIs contain 0.
% As with the the constant drift model results, however, there is some support for increased resistance to PG16 neutralization:  the estimated probability that the mean drift rate is positive is 0.94.
There is little support for any drift rate changes occurring along the phylogeny.
The mean estimated number of rate changes are
0.17 and 0.2 for PG9 and PG16, respectively, and the Bayes factors in favor of relaxed drift
over constant drift are 0.19 and 0.20.
% and the posterior probabilities of zero rate changes are 0.84 and 0.83.
Hence there is not much evidence of localized directional trends that differ
from the overall directional trends.
\par
We illustrate the evolutionary pattern for resistance against VRC01
under the relaxed drift model
in Figure 3.  The branches of
the maximum clade credibility tree are colored according to the inferred branch-specific
mean drift rates, and the
tree tips are labeled with subject identifications.  We obtain a posterior mean estimate of
1.13 rate changes, with a posterior probability of 0.88 for exactly one rate change and
probability greater than 0.99 for at least one rate change.
Furthermore, the Bayes factor in favor of relaxed drift over constant drift is 359.04,
providing very strong support for relaxed drift.  As shown in Table 4, the displacement variance
rate decreases as we move from no drift to a constant drift, and then to relaxed drift.
Figure 3 shows a rate change
occurring at the common ancestral node of samples from subjects P001 and P002.
Apart from
the two branches leading to tips P001 and P002 (which
we refer to as ``branch P001'' and ``branch P002,'' respectively),
the branches have essentially identical
mean drift rates of about 0.15 with 95$\%$ BCI (0.09,0.21).
For the blue-colored branch P001, we have an estimated drift of 2.13
with 95$\%$ BCI (0.06, 4.89), and for the red-colored branch P002 the
estimated drift is -1.18 with 95$\%$ BCI (-4.46, 0.22).  Both estimated
drift rates are drastically different from the parent branch rate, and their
BCIs are also much wider.  If either subject P001 or P002 is deleted from the data set,
we infer a constant, significant drift rate of 0.15 over the entire evolutionary history.
Notably, we do not infer any drift rate changes after deleting either P001 or P002.

\begin{table}[htb]
\begin{center}
\begin{tabular}{lcccc}
%\multicolumn{5}{c}{Table 3: HIV-1 Resistance to Broadly Neutralizing Antibodies} \\
\cline{1-5}
%& \multicolumn{2}{c}{Constant Drift} & \multicolumn{2}{c}{Relaxed Drift} \\
Antibody & \multicolumn{2}{c}{Constant Drift} & \multicolumn{2}{c}{Relaxed Drift} \\
\hline
PG9 & 0.08 & (-0.05, 0.19) & 0.07 & (-0.04, 0.18)  \\
PG16 & 0.10 & (-0.02, 0.20) & 0.11 & (-0.01, 0.24) \\
VRC01 & 0.15 & (0.06, 0.24) & 0.15 & (0.09, 0.21) \\
\hline
\end{tabular}
\end{center}
\caption[HIV-1 resistance to broadly neutralizing antibodies (drift rates)]
{HIV-1 resistance to broadly neutralizing antibodies.  Mean drift rates
under constant and relaxed drift models for
$\log($IC$_{50})$ measurements corresponding to monoclonal neutralizing antibodies PG9,
PG16, and VRC01.  Higher $\log($IC$_{50})$ values represent lower sensitivity to antibody
neutralization, and positive drift rates indicate a trend over time toward greater resistance.
We report posterior means along with 95$\%$ BCIs.
\label{tb:hivresistance}}
\end{table}

\begin{table}[htb]
\begin{center}
\begin{tabular}{lcccccc}
%\multicolumn{5}{c}{Table 4: HIV-1 Resistance to Broadly Neutralizing Antibodies} \\
\cline{1-7}
& \multicolumn{6}{c}{Displacement Variance} \\
Antibody & \multicolumn{2}{c}{No Drift}  & \multicolumn{2}{c}{Constant Drift} & \multicolumn{2}{c}{Relaxed Drift} \\
\hline
PG9 & 0.28 & (0.18, 0.57) & 0.26 & (0.16, 0.53) &  0.26 & (0.17, 0.55) \\
PG16 & 0.26 & (0.17, 0.51) & 0.23 & (0.15, 0.49) &  0.23 & (0.15, 0.51) \\
VRC01 & 0.19 & (0.11, 0.43) & 0.13 & (0.08, 0.27) & 0.06 & (0.04, 0.14) \\
\hline
\end{tabular}
\end{center}
\caption[HIV-1 resistance to broadly neutralizing antibodies (displacement
variance rates)]
{HIV-1 resistance to broadly neutralizing antibodies.  Displacement variance rate
under drift-neutral, constant drift and relaxed drift models for
$\log($IC$_{50})$ measurements corresponding to monoclonal neutralizing antibodies PG9,
PG16, and VRC01.  We report posterior means along with 95$\%$ BCIs.
\label{tb:hivresistance2}}
\end{table}

\begin{figure}
	\centering
	 \includegraphics[width=1.0\textwidth]{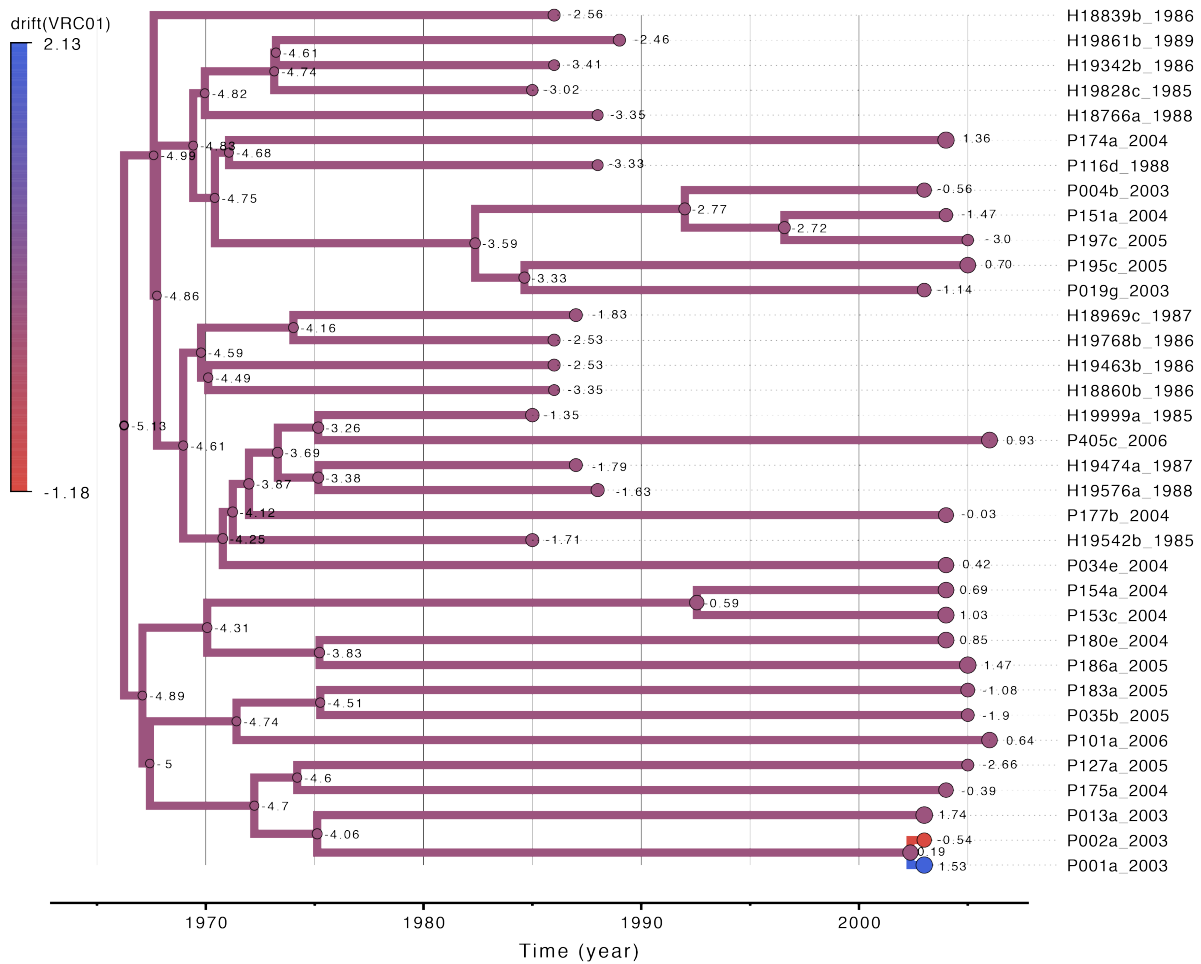}
 	\caption[Maximum clade credibility tree depicting evolutionary pattern of
 	HIV-1 resistance to neutralization by antibody VRC01]
 	{Maximum clade credibility tree depicting evolutionary pattern of
 	HIV-1 resistance to neutralization by antibody VRC01.
 	Subject identifiers corresponding to tree tips
 	are listed to the right of the tree.  The posterior mean drift
 	is depicted using a color gradient along the branches.  Positive drift rates
 	correspond to a trend toward greater resistance to neutralization.
 	The estimated posterior mean drift rate along the purple colored branches that make
 	up most of the tree is 0.15.  A drift rate change is inferred at the common
 	ancestor of tips P001 and P002.
 	The red colored branch leading to the tip P002
 	has an estimated posterior mean drift of -1.18, and
 	the blue colored branch leading tip P001 has an estimated posterior mean
 	drift of 2.13.
 	Tree tips
 	and internal nodes are annotated with observed and inferred $\log($IC$_{50})$
 	values.}
 	\label{Fig. 3}
\end{figure}

\par
Under the relaxed drift model, situations in which both child branches have drift rates
that differ from the drift rate of the parent branch must be handled with care.
In any given tree in the posterior sample,
one of the two branches must inherit its drift rate from the parent branch.
An examination of the posterior distribution of the rate change indicator at the parent node of
tips P001 and P002 reveals that branch P001 inherits the parent branch drift rate with posterior probability of about 40$\%$ and branch P002 inherits the parent rate with the other 60$\%$ probability mass.
Thus the posterior drift rate estimate for each of branches P001 and P002 averages
over the cases where it inherits the parent drift rate and the cases where it differs
from the parent rate.
While the potentially large departures from the parent drift rate still come through
in the ``averaged'' posterior estimates, it is of interest to find out how representative
they are of the true drift rates.  It is conceivable that the ``inherited'' portion of the
posterior may bias the estimate, shifting the mean and widening the BCI.  In the case of
branch P002, for example, the posterior mass near the parent drift rate extends the BCI
into the positive axis so that it includes zero.  It is also conceivable that the
``inherited'' part of the posterior represents a part of the distribution that would
show up even without the restrictions of the relaxed drift model.
\par
To elucidate the true nature of the drift rate change that occurs at the parent
of tips P001 and P002, we conduct a follow-up analysis.  We introduce a new parameterization
of our drift
diffusion model that posits three unique drift rates: one each corresponding to branches
P001 and P002, and another for all remaining branches in the phylogenetic tree.  The results,
presented in Table 5, are similar to the findings from the relaxed drift model.
Notably, the 95$\%$ BCIs for the drift on branches P001 and P002 still contain the range of
credible values for the parent drift rate.
There are also some key differences between the two analyses.
The distributions for the drift on branches
P001 and P002 are bell-shaped rather than bimodal, and the 95$\%$ BCIs are wider than
under the relaxed drift model.  Unlike the relaxed drift model estimate, the 95$\%$ BCI
for the drift on branch P001 contains 0.  However, it has a 0.95 posterior probability
of being positive, so there is still some support that the drift on branch P001
is statistically significant.
\par
An examination of the maximum clade credibility tree in Figure 3,
annotated with observed $\log($IC$_{50})$ values and inferred ancestral trait realizations,
clarifies why we infer a drift rate change.  Consider node triples consisting of two nodes and their common parent node.
The triple of tips P001, P002 and their common ancestor features a relatively large
difference in trait values at the child nodes as well as relatively short branch lengths connecting
nodes P001 and P002 to their parent.  While there are other triples with child nodes possessing
a comparable difference in $\log($IC$_{50})$ values, they have much longer branches leading
from the parent node to the children.  Similarly, while other triples feature relatively
short branches connecting the parent to the children, the trait values at the child nodes
do not differ as much.  The unique combination of short branches coupled with a large difference
between $\log($IC$_{50})$ values at the child nodes explains why the drift rate present on most of
the tree may be incompatible with the triple of P001, P002 and their parent.

\begin{table}[htb]
\begin{center}
\begin{tabular}{lrrrrrr}
%\multicolumn{5}{c}{Table 5: Rate Changes in HIV-1 Drift Toward Resistance to Antibody VRC01} \\
\cline{1-7}
& \multicolumn{3}{c}{Relaxed Drift} & \multicolumn{3}{c}{Fixed-Changes} \\
Branch & Mean  & 95$\%$ BCI  & P(Drift $>$ 0) & Mean & 95$\%$ BCI & P(Drift $>$ 0) \\
\hline
P001 & 2.13 & (0.06, 4.89) & $>$ 0.99 & 2.27 & (-0.36, 5.10) &  0.95 \\
P002 & -1.18 & (-4.46, 0.22) & 0.60 & -1.81 & (-4.61, 1.03) &  0.10 \\
Other & 0.15 & (0.09, 0.21) & $>$ 0.99 & 0.14 & (0.08, 0.20) & $>$ 0.99 \\
\hline
\end{tabular}
\end{center}
\caption[Rate changes in HIV-1 drift toward resistance to antibody VRC01]
{Rate changes in HIV-1 drift toward resistance to antibody VRC01.  We report
posterior means, 95$\%$ BCIs, and posterior probabilities that the drift $>0$.
\label{tb:hivresistance3}}
\end{table}

\par
Although there is strong evidence of a drift rate change at the ancestral node of samples
P001 and P002, the wide BCIs for the branch P001 and branch P002 drift rate estimates suggest that
they are poorly informed by the data.
The type of drift rate change that occurs at the ancestral node of tips P001 and P002 also remains
unclear.  Subjects P001 and P002 are a transmission couple and it appears that one person mounted
a very different antibody response than the other to a highly similar virus.  Further research may clarify the situation.
Nevertheless, we infer a clear, significant drift towards increased resistance to neutralization by
VRC01, and it is robust to deletion of either subject P001 or P002.  At the population level, the
phylogenetic structure of HIV is ``starlike,'' featuring multiple co-circulating lineages,
the dynamics of which generally reflects neutral epidemiological processes
\citep{Grenfell2004}.  It is therefore notable to find evidence
of population level evolution towards increased resistance.

\section{Discussion}

Standard Brownian diffusion is a popular and, in many ways, natural starting point for modeling
continuous trait evolution in a phylogenetic context.  On the other hand, it is very restrictive
and may not adequately describe the dynamics of the underlying evolutionary process.
Development of non-Brownian models, such as mean-reverting Ornstein-Uhlenbeck processes, represents a promising
avenue.  However, 
substantial gains can also be made through 
%more flexible 
building upon standard
Brownian diffusion approaches.
For example, the displacement along a branch is typically assumed to have variance equal to the
product of the branch length and a diffusion variance rate matrix $\textbf P^{-1}$, where
$\textbf P^{-1}$ does not vary along the phylogenetic tree.
\citet{Lemey2010} demonstrate improvements by relaxing this homogeneity assumption via
branch-specific diffusion rate scalars that yield a mixture of Brownian processes.  
Here, we show that progress towards a more realistic 
%Brownian
trait
diffusion can be made by relaxing the assumption of a zero-mean displacement. 
Furthermore, the drift diffusion approach we consider is very general.
Notably, the Ornstein-Uhlenbeck process is nested within the drift diffusion process
defined by 
\begin{equation}
\textbf{Y}_i | \textbf{Y}_{pa(i)} 
\sim N \left(\boldsymbol \beta_1(t_i) \textbf{Y}_{pa(i)} +
 \boldsymbol \beta_2(t_i) \boldsymbol \mu_i, \boldsymbol \Sigma(t_i) \right).
\end{equation}
Consider the special case where $\boldsymbol \mu_i = \boldsymbol \mu$ for every branch, and
\begin{equation}
\boldsymbol \beta_1(t_i) = e^{-\alpha t_i},~~
\boldsymbol \beta_2(t_i) = 1-e^{-\alpha t_i},~~\text{and}~~
\boldsymbol \Sigma(t_i) = \frac{\sigma^2}{2 \alpha} \left[1 - e^{-2\alpha t_i} \right].
\end{equation}
This is equivalent to an Ornstein-Uhlenbeck process on the phylogenetic tree defined by the stochastic differential 
equation
\begin{equation}
d \textbf Y_t = \alpha (\boldsymbol \mu - \textbf Y_t) dt + \sigma d \textbf W_t,
\end{equation}
where $\textbf W_t$ is a standard Brownian diffusion process.  Here, $\boldsymbol \mu$ can be
thought of as an optimal trait value, $\alpha$ represents the strength of selection towards
$\boldsymbol \mu$, and $\sigma^2$ is the variance of the Brownian diffusion component.
Such generality
enables formal testing between a wide class of different Gaussian process models.
\par
We introduce a flexible new Bayesian framework for phylogenetic trait evolution, modeling the
evolutionary process as Brownian diffusion with a nontrivial drift.  By allowing an estimable
mean vector in the displacement distribution, we can account for and quantify a directional
trend.  However, imposing a constant drift rate can make for an unrealistic approximation of the
underlying process.  We overcome this limitation through the relaxed drift model.  The relaxed drift
model permits drift rate variation along a phylogenetic tree while maintaining model identifiability.
Drift rates are generally passed on from parent branches to child branches, and variation is achieved
by allowing at most one branch of any given pair of child branches to assume a different drift rate from their
common parent branch.
\par
The utility of incorporating drift into the diffusion is corroborated by our
analyses of three viral examples.  We apply our methodology to both geographic traits in a phylogeographic
setting as well as phenotypic traits.  Our phylogeographic analysis of the spread of HIV-1 in central Africa
confirms the findings obtained by discrete phylogeographic inference \citep{Faria2014}.  Drift diffusion models fit the data better than
drift-neutral Brownian diffusion, and we uncover directional trends in the dispersal
of the virus from its origin to sampling locations.  We also see that drift rate variation characterizes
real spatiotemporal diffusion processes.  The absence of drift in the diffusion model can lead to
conflation of the latent drift with other parameters, particularly the displacement variance.  Our
analysis of the spread of HIV-1 illustrates how inferred displacement variance rates can decrease with appropriate
drift rate modeling, revealing a clearer, more detailed picture of dispersal dynamics.
\par
While it is tempting to assume drift rate variation and seek out the additional insight it may provide,
the data may not support multiple drift rates.  This may be the case even when there is a significant
constant drift, as we see in our analysis of the West Nile virus.  Parameterizing the model to allow
the maximal number of unique drift rates can result in numerous small rate changes that are a consequence
of the modeling choice and are not necessarily driven by the data.  Our relaxed drift framework
overcomes this issue by inferring the locations and types of rate changes directly from the data
as opposed to making a priori assumptions about the number of unique drift rates and their appropriate
assignments.  Bayesian stochastic search variable selection enables efficient exploration of all
possible drift rate configurations.
\par
Although our focus has been on drift, a major strength of our approach is its implementation
in the larger Bayesian phylogenetic framework of BEAST.
Through BEAST, we have access to a
plethora of different models for molecular character substitution, demographic history and
molecular clocks.
Bayesian inference provides a natural
framework for controlling for different sources of uncertainty in evolutionary models,
including the phylogenetic tree and trait and sequence evolution parameters, and testing evolutionary
hypotheses.
\par
The gains from introducing drift into our real data examples are encouraging, and there is a
need for continued development of more realistic trait evolutionary models.  We anticipate that our
drift diffusion approach will be useful in other scenarios not examined here, including antigenic
drift in influenza.  Antigenic drift is the process by which influenza viruses evolve to evade the
immune system, and an understanding of its dynamics is essential to public health efforts.
\citet{Bedford2014} have recently developed an integrated approach to mapping antigenic phenotypes that
%GB: 'combines it' sounds strange to me
combines it with genetic information.  It may be fruitful to model the diffusion of the antigenic phenotype
in their framework with a relaxed drift.
\par
While the relaxed drift model has proven to be flexible and useful, its identifiability restrictions
may render it inappropriate for some evolutionary scenarios.  For example, once
a drift rate appears anywhere in the phylogenetic tree, the restrictions mandate that it must be passed on
and ``survive'' until it reaches an external branch.  Lack of support for the
survival of a specific drift rate may not preclude its inclusion under relaxed drift.
In the analysis of HIV-1 resistance to neutralization, for example, the parent branch of branches
P001 and P002 must pass on its drift rate to one of the two child branches in each tree in the
posterior sample.  Yet, the branch which is forced to inherit the parent drift rate alternates between P001
and P002 in the posterior sample, resulting in posterior drift distributions for both branches that
differ from that of their parent drift.
However, this unnatural mechanism of deflecting an unsupported drift rate may contribute to
misleading drift rate estimates.  %PL: isn't a bit strange in this respect that variance appears to be higher for the fixed-changes approach you tried?
It would be preferable to sidestep such problems by developing
alternative models that accommodate drift rate variation while retaining identifiability.

\section*{Acknowledgements}
The research leading to these results has received funding from the European Research Council
under the European Community's Seventh Framework Programme (FP7/2007-2013) under
Grant Agreement no.~278433-PREDEMICS and ERC Grant agreement no.~260864 and the
National Institutes of Health (R01 AI107034, R01 HG006139, R01 LM011827 and T32 AI007370) and the
National Science Foundation (DMS 1264153).

\appendix

\section*{Appendix}
\section{Identifiability}

%\section{Appendix: Identifiability}

To ensure that our results are meaningful, it is important to understand the conditions
under which our model is identifiable.
For convenience, and without loss of generality, we assume here that traits are one-dimensional.
Following the development in section 2.1,
the observed traits $\textbf Y = (\textbf Y_1 ,\dots,\textbf Y_N)^t$ at the tips of the phylogeny $\tau$
are multivariate-normal distributed:
\begin{equation}
P \left(\textbf{Y}  \vert \textbf{Y}_{2N-1}, \textbf{P}, \textbf{V}_{\tau}, \boldsymbol \mu_{\tau} \right) =
\mbox{MVN} \left(\textbf{Y}; \textbf{Y}_{\text{root}} + \textbf{T}\boldsymbol \mu_{\tau}, \textbf{P}^{-1} \otimes
\textbf{V}_{\tau} \right).
\end{equation}
Here, $\textbf Y_{\text{root}}$ is
an $N \times 1$ vector with the root trait $\textbf Y_{2N-1}$ repeated $N$ times.
The
vector $\boldsymbol \mu_{\tau} = (\boldsymbol \mu_{1},\dots,\boldsymbol \mu_{2N-2})^{t}$
consists of branch-specific drift rates $\boldsymbol \mu_i$.
The $N \times N$ variance matrix $\textbf{V}_{\tau}$ is a deterministic function of $\tau$
and represents the contribution of the phylogenetic tree to the covariance structure.
Its diagonal entries $V_{ii}$ are equal to the distance in time between the tip $\mathcal{V}_i$ and the
root node $\mathcal{V}_{2N-1}$, and off-diagonal entries $V_{ij}$ correspond to the distance in time between the
root node $\mathcal{V}_{2N-1}$ and the most recent common ancestor of tips $\mathcal{V}_{i}$ and $\mathcal{V}_{j}$.
Finally, the $N \times (2N-2)$ matrix $\textbf T$ is defined as follows: $T_{ij} = t_j$,
the length of branch $j$, if branch
$j$ is part of the path from the external node $i$ to the root, and $T_{ij} = 0$ otherwise.  In other words,
the $i$th row of $\textbf T$ specifies the path of branches in $\tau$ connecting external node $i$ to
the root.
\par
Let $\textbf V(\boldsymbol \mu_{\tau})$ denote the vector space of permissible values of $\boldsymbol \mu_{\tau}$
for our model.  With respect to the drift,
the model is identifiable if the equality
\begin{equation}
P \left(\textbf{Y}  \vert \textbf{Y}_{2N-1}, \textbf{P}, \textbf{V}_{\tau}, \boldsymbol \mu_{\tau} \right) =
P \left(\textbf{Y}  \vert \textbf{Y}_{2N-1}, \textbf{P}, \textbf{V}_{\tau}, \boldsymbol \mu_{\tau}^{*} \right)
\end{equation}
implies that $\boldsymbol \mu_{\tau} = \boldsymbol \mu_{\tau}^{*}$.  Because the drift appears only in the mean
of the distribution,
we have an identifiability problem if the same mean $\mathrm{E}(\textbf Y)$ can be realized from
different values of $\boldsymbol \mu_{\tau}$.  In other words, if there exist
$\boldsymbol \mu_{\tau} \neq \boldsymbol \mu_{\tau}^{*}$ such that $\textbf T \boldsymbol \mu_{\tau} =
\textbf T \boldsymbol \mu_{\tau}^{*} $.  This can happen if and only if the
linear transformation $\textbf T$ has a nontrivial kernel.  We know that
\begin{equation}
\mbox{dim} \textbf V (\boldsymbol \mu_{\tau}) = \mbox{dim ker}(\textbf T) +
\mbox{dim range} (\textbf T) ,
\end{equation}
and we also know that for any phylogeny $ \tau$, $\textbf T$ is of full rank
because its rows are linearly independent.
It follows that for the kernel of $\textbf T$ to be trivial, we must have
\begin{equation}
\mbox{dim} \textbf V (\boldsymbol \mu_{\tau}) \leq N .
\end{equation}
If we allow a unique drift rate on each branch of $\tau$, we have
$\textbf V (\boldsymbol \mu_{\tau}) = \mathbb{R}^{2 \mathbb{N} -2}$.  Therefore we must take a different approach.
\par
It is illuminating to look at identifiability from the perspective of linear equations.
For a given $\boldsymbol \mu_{\tau}$, $\textbf T$ maps $\boldsymbol \mu_{\tau}$ to an $N \times 1$ vector
$\boldsymbol \gamma$:
\begin{equation}
\textbf T \boldsymbol \mu_{\tau} = \boldsymbol \gamma .
\end{equation}
Identifiability is then equivalent to the system (52) of $N$ linear equations
having a unique solution.

%Each sum in $\textbf T \boldsymbol \mu_{\tau}$ is defined
%by the path from the root of $\tau$ to one of the $N$ tips of
%$\tau$.
%The root is connected to a tip $i$ by a series of branches, say
%with indices $i1,i2,\dots,i \alpha_j$.  With each branch, we associate a
%product $\textbf t_{ij} \boldsymbol \mu_{ij}$
%of the length of the branch (measured in time) and the drift rate on the branch. Therefore,
%the $ith$ entry of the vector $\textbf T \boldsymbol \mu $ can be written
%\begin{equation}
%\sum_{j=1}^{\alpha_i} \textbf t_{ij} \boldsymbol \mu_{ij},
%\end{equation}
%and the system $\textbf T \boldsymbol \mu = \boldsymbol \gamma$ can be written
%\begin{eqnarray}
%\sum_{j=1}^{\alpha_1} \textbf t_{1j} \boldsymbol \mu_{1j} & = & \boldsymbol \gamma_1 \\
%\sum_{j=1}^{\alpha_2} \textbf t_{2j} \boldsymbol \mu_{2j} & = & \boldsymbol \gamma_2 \\
%\dots \\
%\sum_{j=1}^{\alpha_N} \textbf t_{Nj} \boldsymbol \mu_{Nj} & = & \boldsymbol \gamma_N .
%\end{eqnarray}
%\par
To achieve identifiability, we introduce the relaxed drift model.  Starting with a drift rate
on the unobserved branch leading to the root node and moving down the tree toward the external
nodes, every time a branch splits into two branches,
one of two things happens.  
Either both of the child branches inherit the drift rate of the
parent branch, or exactly one of the child branches inherits the drift rate from the
parent branch while the other gets a new drift rate.  Both child
branches taking on different drift rates than the parent branch is not permitted.
To avoid confusion, we continue to denote the $2N-2$ branch-specific
drift rates as $\boldsymbol \mu_1,\dots,\boldsymbol \mu_{2N-2}$, with the understanding that they are not all unique.
We let $\boldsymbol \mu^{*}_1,\dots,\boldsymbol \mu^{*}_K$ denote the unique drift rates, where
$K \leq N$.  \\[1ex]
%GB: consistent notation: T_{i,x} ?
\textbf{Definition:} We say that a row $\textbf T_i = (T_{i1}, T_{i2},\dots,T_{i,2N-2})$ in
$\textbf T$ is $\boldsymbol \mu^{*}_k -\textit{dominated}$ if its associated path from the root
of $\tau$ to a tip ends
with a branch with drift rate $\boldsymbol \mu^{*}_k$.  We also refer to the
sum $\textbf T_i \boldsymbol \mu_{\tau} = \sum_{j=1}^{2N-2} T_{ij} \boldsymbol \mu_j$ and
the path associated with $\textbf T_i$
as $\boldsymbol \mu^{*}_k -\textit{dominated}$. \\[1ex]
Note that each unique drift rate dominates at least one path. \\[1ex]
\textbf{Definition:} An \textit{initial branch} of the rate $\boldsymbol \mu^{*}_k$
is a branch whose parent branch has a different drift rate.  The unobserved branch leading to
the root node is also defined to be an initial branch. \\[1ex]
Observe that every branch with drift $\boldsymbol \mu^{*}_k$ is an initial branch of
$\boldsymbol \mu^{*}_k$ or a descendant of an initial branch
of $\boldsymbol \mu^{*}_k$.  A drift rate may have more than one initial branch.
In order to quantify how deep into the tree $\tau$ a drift rate extends
(starting from the tips and going toward the root), we make the following definition. \\[1ex]
\textbf{Definition:} By a \textit{descendant path} of a branch $b$, we mean a
series of connected branches, starting with a child branch of $b$
and ending with a branch leading to a tip.
The \textit{depth} of a branch $b$ is equal to the maximal number of branches in descendant paths
of $b$.  The depth of a drift rate $\boldsymbol \mu^{*}_k$ is equal to the
maximal depth of its initial branches. \\[1ex]
For example, if $\boldsymbol \mu^{*}_k$ has one initial branch leading to
a tip, then $\boldsymbol \mu^{*}_k$ has depth 0.  If the number of unknowns $K$ is less than the number of
equations $N$ in the system $\textbf T \boldsymbol \mu_{\tau} = \boldsymbol \gamma$, a unique solution
can be established by working with a reduced system.  We form the reduced system by choosing $K$
of the $N$ rows in $\textbf T$, say $\textbf T_{i_1}, \dots, \textbf T_{i_K}$,
such that each is dominated by a different drift rate.  If a drift rate dominates more than one path,
we choose a path containing a maximal depth initial branch of the rate for the reduced system. \\[1ex]
%Let $\mathcal R$ denote the reduced set of branches, meaning branches that appear in at least one path associated with a row in the reduced set of $K$ rows.
%\textbf{Claim:} Consider a $\boldsymbol \mu^_k-$dominated path $P$ that contains the initial branch
%of $\boldsymbol \mu^{*}_j$, $j \neq k$.  Then the depth of $\boldsymbol \mu^{*}_k$
%is less than the depth of $\boldsymbol \mu^{*}_j$.  \\[1ex]
%\textbf{Proof:} $P$ contains both the initial branch of $\boldsymbol \mu^{*}_j$
%and the initial branch of $\boldsymbol \mu^{*}_k$, and since $P$ is
%$\boldsymbol \mu^{*}_k-$dominated, the initial branch of $\boldsymbol \mu^{*}_k$
%is a descendant of $\boldsymbol \mu^{*}_i$
\textbf{Claim:} The relaxed drift model is identifiable.  \\[1ex]
\textbf{Proof:} The reduced linear system
%GB: no need to number all the lines below? It messes up referencing to this set of equations later on. Use \nonumber
\begin{eqnarray}
\sum_{j=1}^{2N-2} T_{i_1j} \boldsymbol \mu_{j} & = & \boldsymbol \gamma_{i_1} \\
\sum_{j=1}^{2N-2} T_{i_2j} \boldsymbol \mu_{j} & = & \boldsymbol \gamma_{i_2} \\
\dots \\
\sum_{j=1}^{2N-2} T_{i_Kj} \boldsymbol \mu_{j} & = & \boldsymbol \gamma_{i_K},
\end{eqnarray}
consists of $K$ equations and $K$ variables.
Therefore to show that the solution is unique, it suffices to show that the
linear system is independent.  To establish independence, it suffices to show that if
\begin{equation}
a_1 \sum_{j=1}^{2N-2} T_{i_1j} \boldsymbol \mu_j
+ a_2 \sum_{j=1}^{2N-2} T_{i_2j} \boldsymbol \mu_j
\dots + a_N \sum_{j=1}^{2N-2} T_{i_Kj} \boldsymbol \mu_j
= 0  ,
\end{equation}
for some constants $a_1,\dots,a_K$, then we must have
\begin{equation}
a_1 = a_2 = \dots = a_K = 0.
\end{equation}
Suppose (57) holds.  The idea behind the proof is as follows: we consider all
drift rates of depth 0, conclude that each
sum in (57) dominated by a drift rate of depth 0 must have its corresponding
coefficient $a_i = 0$, then consider all drift rates of depth 1,
conclude that each
sum in (57) dominated by a drift rate of depth 1
must have corresponding coefficient $a_i = 0$,  and so on until we
have gone through all possible depth values
of drift rates in $\tau$.
\par
Suppose $\boldsymbol \mu^{*}_k$ has depth 0.  Then $\boldsymbol \mu^{*}_k$ only appears in the
single $\boldsymbol \mu^{*}_k-$dominated sum and cannot be canceled out by a linear combination of
the other sums.  This forces the coefficient $a_i$ of the $\boldsymbol \mu^{*}_k-$dominated
sum in (57) to be equal to zero.  Having shown that any sum dominated by a drift rate of depth
0 must have a zero coefficient in (57), we can move on to the case of depth 1.
%If $\boldsymbol \mu^{*}_k$ has depth 1, then it appears in the
%$\boldsymbol \mu^{*}_k-$dominated path, and it may appear in paths dominated by other rates.
%Suppose $\boldsymbol \mu^{*}_k$ appears in a path $P_i$ dominated by a different rate, say
%$\boldsymbol \mu^{*}_i$.  By construction of the reduced system, $P_i$ contains an initial branch
%$\textbf b_i$ of $\boldsymbol \mu^{*}_i$ of maximal depth.  This means the depth of $\textbf b_i$
%is equal to the depth of $\boldsymbol \mu^{*}_i$.  Because $\textbf b_i$ is a descendant of a branch with
%rate $\boldsymbol \mu^{*}_k$, the depth of $\boldsymbol \mu^{*}_i$ must be less than the depth of
%$\boldsymbol \mu^{*}_k$.  But sums dominated by rates of depth less than 1 have already been shown to have zero
%coefficients in (56).  Thus the sum associated with $P_i$ has coefficient zero.  Because
%$\boldsymbol \mu^{*}_k$ appears in only one sum which is not already known to have a zero coefficient,
%the $\boldsymbol \mu^{*}_k-$dominated sum, it follows that in order for (56) to hold, the
%$\boldsymbol \mu^{*}_k-$dominated sum must also have a zero coefficient.
%\par
Rather than handle the case of depth 1 separately, we present a general argument.
\par
Suppose the coefficients of all
sums in (57) that are dominated by drift rates of depth less than $m$ have been shown to be zero.
Consider drift rates of depth $m$.
If $\boldsymbol \mu^{*}_k$ has depth $m$, then it appears in the
$\boldsymbol \mu^{*}_k-$dominated path, and it may appear in paths dominated by other rates.
Suppose $\boldsymbol \mu^{*}_k$ appears in a path $P_i$ dominated by a different rate, say
$\boldsymbol \mu^{*}_i$.  By construction of the reduced system, $P_i$ contains an initial branch
$\textbf b_i$ of $\boldsymbol \mu^{*}_i$ of maximal depth.  This means the depth of $\textbf b_i$
is equal to the depth of $\boldsymbol \mu^{*}_i$.  Because $\textbf b_i$ is a descendant of a branch with
rate $\boldsymbol \mu^{*}_k$, the depth of $\boldsymbol \mu^{*}_i$ must be less than the depth of
$\boldsymbol \mu^{*}_k$.  But sums dominated by rates of depth less than $m$ have already been shown to have zero
coefficients in (57).  Thus the sum associated with $P_i$ has coefficient zero.  Because
$\boldsymbol \mu^{*}_k$ appears in only one sum which is not already known to have a zero coefficient,
the $\boldsymbol \mu^{*}_k-$dominated sum, it follows that in order for (57) to hold, the
$\boldsymbol \mu^{*}_k-$dominated sum must also have a zero coefficient.  Therefore sums dominated by drift
rates of depth $m$ must have zero coefficients in (57).
\par
Invoking this argument until we have gone through all possible values of drift rate depth,
it follows that $a_1 = a_2 = \dots = a_K = 0$. $ \blacksquare $

\clearpage

\bibliographystyle{sysbio}

\bibliography{list_of_references}

\end{document}